\documentclass[superscriptaddress,reprint,amsmath,amssymb,aps,pra,floatfix]{revtex4-1}
\usepackage[margin=1.5cm]{geometry}

\usepackage{xcolor} 
\usepackage{graphicx}
\usepackage{dcolumn}
\usepackage{upgreek}
\usepackage{bm}
\usepackage{hyperref}
\usepackage{floatrow}
\usepackage{times}
\usepackage[mathlines]{lineno}
\usepackage{physics}
\usepackage{color,soul}

\usepackage{etoolbox}


\begin{document}

\title{A low-loss ferrite circulator as a tunable chiral quantum system}
\author{Ying-Ying Wang}
\affiliation{Department of Physics, University of Massachusetts-Amherst, Amherst, MA, USA}
\author{Sean van Geldern}
\affiliation{Department of Physics, University of Massachusetts-Amherst, Amherst, MA, USA}
\author{Thomas Connolly}
\email{Present address: Department of Applied Physics, Yale University, New Haven, CT, USA}
\affiliation{Department of Physics, University of Massachusetts-Amherst, Amherst, MA, USA}
\author{Yu-Xin Wang}
\affiliation{Pritzker School of Molecular Engineering, University of Chicago, Chicago, IL, USA}
\author{Alexander Shilcusky}
\affiliation{Department of Physics, University of Massachusetts-Amherst, Amherst, MA, USA}
\author{Alexander McDonald}
\affiliation{Pritzker School of Molecular Engineering, University of Chicago, Chicago, IL, USA}
\affiliation{Department of Physics, University of Chicago, Chicago, IL, USA}
\author{Aashish A. Clerk}
\affiliation{Pritzker School of Molecular Engineering, University of Chicago, Chicago, IL, USA}
\author{Chen Wang}
\email{wangc@umass.edu}
\affiliation{Department of Physics, University of Massachusetts-Amherst, Amherst, MA, USA}

\date{\today}

\begin{abstract}

Ferrite microwave circulators allow one to control the directional flow of microwave signals and noise, and thus play a crucial role in present-day superconducting quantum technology.  They are typically viewed as a black-box, with their internal structure neither specified nor used as a quantum resource.
In this work, we show a low-loss waveguide circulator constructed with single-crystalline yttrium iron garnet (YIG) in a 3D cavity, and analyze it as a multi-mode hybrid quantum system with coupled photonic and magnonic excitations. 
We show the coherent coupling of its chiral internal modes with integrated superconducting niobium cavities, and how this enables tunable non-reciprocal interactions between the intra-cavity photons.  We also probe experimentally the effective non-Hermitian dynamics of this system and its effective non-reciprocal eigenmodes.  The device platform provides a test bed for implementing non-reciprocal interactions in open-system circuit QED.

\end{abstract}

\maketitle

\section{Introduction}

Microwave circulators, typically composed of a transmission line Y-junction with ferrite materials~\cite{kord_microwave_2020}, are ubiquitous in superconducting circuit QED experiments~\cite{devoret_superconducting_2013}.  They provide a crucial link in the readout chain of superconducting quantum processors, by directing the signal traffic while protecting the qubits and resonators from thermal noise~\cite{krantz_quantum_2019}.  They also 
enable the interactions between distinct quantum circuit modules to be non-reciprocal~\cite{kurpiers_deterministic_2018, axline_-demand_2018}, a feature which is important for eliminating long-distance cross-talks in modular quantum computation architectures.  Despite their importance, microwave circulators are generally treated as broadband black-box devices in experiments.  Formulating a more microscopic quantum description is often challenging, as their internal modes involving the magnetic spin excitations (magnons) are generally too lossy and complex to be analyzed using canonical circuit quantization~\cite{vool_introduction_2017}.

On the other hand, there has been growing interest in studying and manipulating magnon excitations of ferromagnetic/ferrimagnetic materials in the quantum regime~\cite{lachance-quirion_hybrid_2019, awschalom_quantum_2021}.  In particular, the ferromagnetic resonance (FMR) mode of yttrium iron garnet (YIG), a ferrimagnetic insulator with usage in commercial circulators, has shown sufficiently high quality factor and coupling cooperativity with microwave cavities to function as a quantum oscillator mode in strong-coupling circuit QED~\cite{huebl_high_2013,tabuchi_hybridizing_2014, zhang_strongly_2014}.  Notably, coherent coupling of magnons with a superconducting qubit~\cite{tabuchi_coherent_2015} and single-shot detection of a single magnon~\cite{lachance-quirion_entanglement-based_2020} have been demonstrated using a millimeter-sized single-crystalline YIG sphere in a 3D cavity. 
Furthermore, there is a plausible pathway towards planar superconducting-magnonic devices~\cite{hou_strong_2019, golovchanskiy_ultrastrong_2021} to connect circuit QED with spintronics technologies by advancing fabrication techniques of low-damping YIG films~\cite{heyroth_monocrystalline_2019}. 

It would be interesting to harness these recent advances in the study of quantum magnonics to revisit the design of microwave circulators, potentially leading to new kinds of non-reciprocal devices in circuit QED. Our work here describes a first step in this direction.  Here we demonstrate a tunable non-reciprocal device based on the waveguide circulator loaded with single-crystalline YIG, which explicitly makes use of well-characterized hybrid polariton modes.  Such modes are the normal modes of coupled magnon-photon systems~\cite{huebl_high_2013,tabuchi_hybridizing_2014, zhang_strongly_2014, boventer_complex_2018, zhang_observation_2017}, and have an intrinsic chirality that is set by the magnetic field~\cite{anderson_engineering_2016, owens_quarter-flux_2018, zhang_broadband_2020}.
While our device follows the same basic working principles underpinning textbook circulators~\cite{kord_microwave_2020, fay_operation_1965},
detailed understanding of the internal modes allows us to incorporate the physical source of non-reciprocity in the full description of a larger system including two external superconducting cavities, using a non-Hermitian effective Hamiltonian.

While our device can be configured to operate as a traditional circulator for its non-reciprocal transmission of travelling waves, the main focus of our study is to use the device for mediating tunable non-reciprocal interaction between localized long-lived quantum modes.  Such non-reciprocal mode-mode couplings result in distinct signatures in the eigenvalues and eigenvectors of the non-Hermitian system Hamiltonian, which is relevant to the more general study of non-Hermitian dynamics in contexts ranging from classical optics to quantum condensed matter.  
Anomalous properties of the eigenvalues and eigenvectors of a non-Hermitian Hamiltonian have given rise to a number of striking phenomena such as the existence of exceptional points~\cite{heiss_physics_2012, ozdemir_paritytime_2019} and the non-Hermitian skin effect~\cite{hatano_vortex_1997, yao_edge_2018, mcdonald_phase-dependent_2018}, but direct experimental access to the underlying eigenmodes is often difficult.  
In this study, we provide comprehensive characterization of the eigenmode structure, which is a step towards effective Hamiltonian engineering of non-reciprocal non-Hermitian systems. 

The most tantalizing usage of non-reciprocity in quantum systems (such as entanglement stabilization using directional interactions in chiral quantum optics setups~\cite{stannigel_driven-dissipative_2012,lodahl_chiral_2017}) require extremely high quality devices.  In particular, they must approach the pristine limit where undesirable internal loss rates are negligible compared to the non-reciprocal coupling rates.  While many experiments have focused on new avenues of achieving non-reciprocity~\cite{chapman_widely_2017, lecocq_nonreciprocal_2017, sliwa_reconfigurable_2015, ruesink_nonreciprocity_2016, fang_generalized_2017, wang_nonreciprocity_2019, xu_nonreciprocal_2019}, this loss-to-coupling ratio, which can be understood as the quantum efficiency of the non-reciprocal interactions, has been typically limited to approximately 10\% ($\sim$ 0.5 dB) or more, which is comparable to the linear insertion loss of typical commercial circulators as measured in modular circuit QED experiments~\cite{kurpiers_deterministic_2018, axline_-demand_2018}.  This performance lags far behind the quality of unitary operations between reciprocally coupled quantum components (i.e.~two-qubit gate infidelity $<1$\%).  Our approach provides a route for transcending this limitation on the quantum efficiency of non-reciprocal interactions.    

The results of our study have implications in several areas:
(1) In the context of quantum magnonics, we present the first study of polariton modes with a partially magnetized ferrite material, which features a high quality factor and low operating field, both of which are crucial for constructing superconducting-magnonic devices.
(2) In the context of modular superconducting quantum computing, we demonstrate the first circulator with internal loss well below 1\% of the coupling bandwidth, which would enable high-fidelity directional quantum state transfer. 
(3) For the general non-Hermitian physics, we demonstrate an experimental probe of the non-reciprocal eigenvector composition of a non-Hermitian system.  
Combining these advances, we have established an experimental platform that meets the conditions for future study of nonlinear non-reciprocal interactions with superconducting qubits.

\section{Experimental Setup}

\begin{figure}[tbp]
    \centering
    \includegraphics[scale=.36]{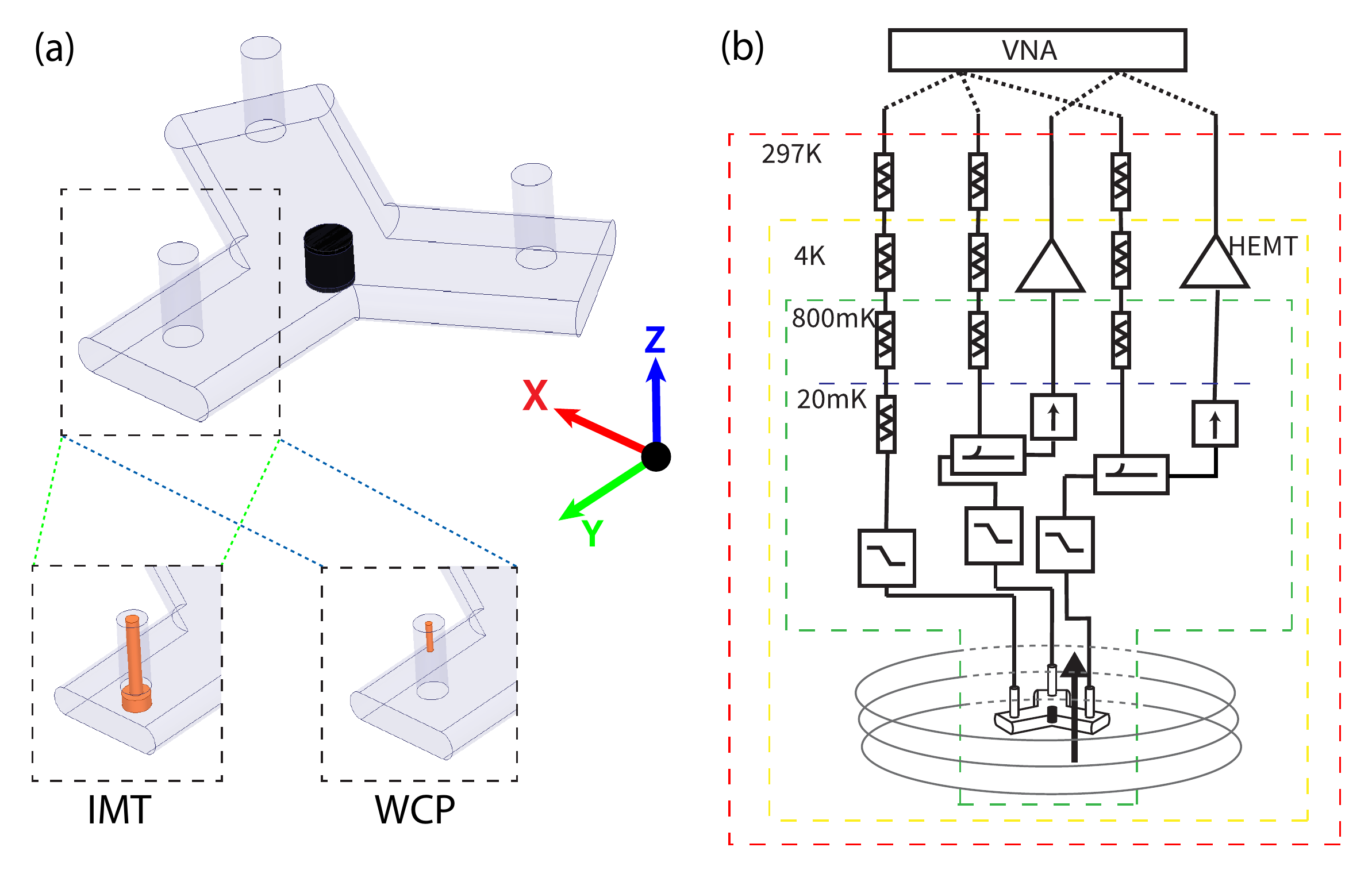}
    \caption{\textbf{Device and measurement setup.} 
    (a) A YIG cylinder (black) is placed at the center of the intersection of three rounded-rectangular waveguides placed 120 degrees away from each other.  The light grey region is vacuum inside an oxygen-free copper enclosure. The device can be assembled in two different configurations: First, a drum-head shaped transition pin can be attached at the end of each waveguide section to form an impedance matched waveguide-to-SMA transition (IMT). Alternatively, a short weakly-coupled probe (WCP) can be attached to each waveguide section to explore the internal modes of the device. (b) The device is mounted to a mezzanine plate that is thermalized to the mixing chamber of a dilution refrigerator, and is positioned at the center of a superconducting solenoid magnet which operates at 4K.  The device is connected to three input cables (with attenuators as marked) and two output amplifier lines (with directional couplers splitting signals) for $S$-parameter measurements using a vector network analyzer (VNA). 
    }
    \label{fig:setup}
\end{figure}

Our experimental setup is shown in Fig.~1(a).
Three 
rounded-rectangular waveguides, each with a cross section of 21.0 mm $\times$ 4.0 mm, placed 120 degrees away from each other, intersect to form the body of the circulator. A $\phi$-5.58 mm $\times$ 5.0 mm single-crystalline YIG cylinder is placed at the center of the Y-junction, with external magnetic fields applied along its height (the $z$ axis and the [111] orientation of the YIG crystal).  
At the end of the three waveguide sections, we can either attach impedance-matched waveguide-to-SMA transitions (IMT) to perform standard characterization of the circulator (as in Section IV), or attach weakly-coupled probe pins (WCP) to explore the internal modes of this YIG-loaded Y-shaped cavity (as in Section III).  The use of reconfigurable probes in the same waveguide package allows us to infer the operation condition and the performance of the circulator from the properties of the internal modes.  Furthermore, the copper waveguide sections can be replaced by superconducting niobium cavities, with details to be described in Section V and Fig.~5.  This modular substitution introduces additional external high Q modes to the system, and understanding the resulting Hamiltonian and the hybridized mode structure of the full system will be a first step towards the study of pristine non-reciprocal interactions in circuit QED.

The device package is thermalized to the mixing chamber plate ($\sim$20 mK) of a Bluefors LD-250 dilution refrigerator inside the $\phi$-100 mm bore of a 1 T superconducting magnet that applies magnetic field along the $z$ axis [Fig.~1(b)].  A vector network analyzer is used to measure the complex microwave transmission coefficients $S_{ij}$ (from Port $j$ to Port $i$, where $i,j=1,2,3$) of the device in series with a chain of attenuators, filters and amplifiers as in typical circuit QED experiments.  A magnetic shield made of a steel sheet is placed outside the bottom half of the refrigerator, and all data is acquired under the persistent mode of the superconducting magnet to minimize magnetic-field fluctuations.

\section{Internal mode structure}
\begin{figure*}[tbp]
    \centering
    \includegraphics[scale=.235]{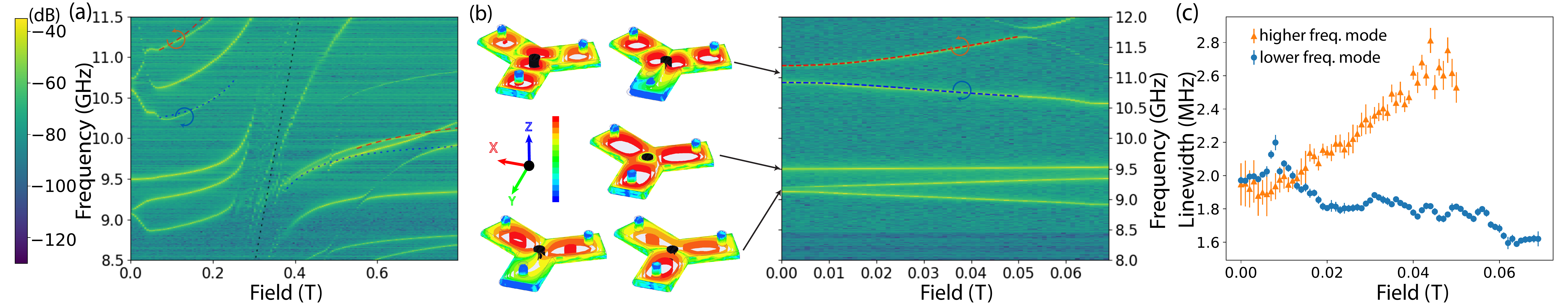}
    \caption{\textbf{Internal mode spectrum of the device.}  (a) VNA transmission measurement $S_{21}$ of multi-mode photon-magnon hybrid system formed in the waveguide circulator package 
     with WCP. The blue (red) dashed line plots the frequency of the clockwise(counterclockwise) mode from a simplified two-mode model of photon-magnon avoided crossing with $g/2\pi$ = 1.3 GHz (2.1 GHz) to compare with an observed spectral line.  (b) The right panel shows a finer sweep of $S_{21}$ in the low-field regime.  The mode frequencies differ slightly from (a) since the data was acquired after some modifications to the device packaging (a piece of Teflon spacer at the top of the YIG cylinder was removed).  The left panel shows the electromagnetic mode structures of the eigenmode solutions from our HFSS simulation for the WCP with good frequency agreement to the experimental data (see Fig.~\ref{fig:eigenfrequency} in Appendix).  The color scale from red to blue represents electric field strength from high to low in log scale.  The pair of modes around 11 GHz are connected to the circulating modes of the loaded circulator and (c) shows their linewidths. 
    \label{fig:internal_spectrum}
    }
\end{figure*}

We begin by discussing the internal mode structure of the device, as probed by $S_{21}$ as a function of applied magnetic field $B$ when the device is installed with 
WCP [Fig.~\ref{fig:internal_spectrum}(a)]
. 
A series of electromagnetic modes (relatively field-independent) are observed to undergo large avoided crossings with a cluster of magnon modes of the YIG crystal, forming photon-magnon polariton modes.  
The magnon mode most strongly coupled to photons is known to correspond to near-uniform precession of YIG spins, or the Kittel mode of FMR, whose frequency increases linearly with magnetic field: $\omega_m=\gamma [B+\mu_0(N_{x,y}-N_z)M_s]\approx \gamma B$, as marked by the dashed line in Fig.~\ref{fig:internal_spectrum}(a).  Here $\gamma$ is the gyromagnetic ratio, and the (volume-averaged) demagnetizing factors $N_{x,y,z}$ in magnetic saturated state are very close to $1/3$ for the aspect ratio of our YIG cylinder~\cite{chen_demagnetizing_1991}.  These avoided crossings are similar to previous experiments showing strong photon-magnon coupling~\cite{tabuchi_hybridizing_2014,zhang_strongly_2014}, but due to the much larger size of the YIG in our experiment, a large cluster of higher-order magnetostatic modes, most of which have slightly higher frequency than the Kittel mode~\cite{fletcher_electron_1960, klingler_gilbert_2017} also coherently interact with the microwave photons, contributing to the complex transmission spectra in the vicinity of the crossings.  Nevertheless, to have a coarse estimate of the photon-magnon coupling strength, it is convenient to model each observed spectral line far away from the crossing region as a bare electromagnetic mode with frequency $\omega_c/2\pi$ hybridized with a single combined magnon mode.
The implied coupling strengths $g/2\pi$ (in the cavity QED convention) are about 1.2 GHz and 2.1 GHz for the two modes of particular interest to this study [blue and red in Fig.~\ref{fig:internal_spectrum}(a)], placing the mode hybridization in the ultrastrong coupling regime 
(see e.g.~\cite{markovic_demonstration_2018}) with $g/(\omega_c+\omega_m) \sim 10\%$.  Even at $B=0$, with a photon-magnon detuning of $\Delta=\omega_c-\omega_m\approx 2\pi\cdot10$ GHz, the participation of magnon excitations in the photon-branch of the polariton modes remains quite substantial.

Using finite-element simulations (Ansys HFSS, Appendix A), we identify that the five polariton modes in the frequency range of 8-12 GHz at $B=0$ include two nearly-degenerate mode pairs with two-fold symmetry and another mode with three-fold symmetry.  Electric field distributions of each of the modes are illustrated in Fig.~\ref{fig:internal_spectrum}(b).  Each degenerate mode pair can be understood using a basis of standing-wave modes polarized along the $x$ or $y$ direction.  The application of a magnetic field lifts this $x$-$y$ degeneracy, as the mode pair forms clockwise and counterclockwise rotating eigenmodes with a frequency splitting~\cite{owens_quarter-flux_2018, zhang_broadband_2020, anderson_engineering_2016}. 

Prior use of these chiral polariton mode pairs have been in the magnetically saturated regime~\cite{anderson_engineering_2016, owens_quarter-flux_2018, zhang_broadband_2020}.  Here we focus on the low-field regime ($|B|<0.05$ mT) where the approximately linear increase of frequency splitting between the mode pair reflects increasing magnetization of YIG under increasing applied magnetic field.  After implementing demagnetization training cycles to suppress a relatively small hysteretic effect throughout our experiments, we expect an approximately linear magnetization curve ($M$-$H$) for YIG until it approaches magnetic saturation.  In the limit of high permeability $\mu\gg \mu_0$
(with $\mu_0$ being the vacuum permeability), we have $M=B/{\mu_0 N_z}$ (note that $B$ is the applied magnetic field strength) and $N_z\approx 0.285$ is the $z$-direction demagnetizing factor when the YIG is significantly below magnetic saturation~\cite{chen_demagnetizing_1991}.  Saturation magnetization $M_s=2440$ Oe~\cite{solt_temperature_1962} of YIG is approached on the scale of $B\sim \mu_0N_zM_s\approx70$ mT, which agrees with the changing curvature of the mode-splitting spectra. 

On the other hand, in the completely demagnetized state ($M=0$) at zero field, the system is expected to satisfy macroscopic time-reversal symmetry.  As supported by numerical simulations, the $x$-$y$ mode pairs should be in principle exactly degenerate since both the Y-junction geometry and the [111] YIG crystal has 3-fold rotational symmetry around the $z$ axis.
However, appreciable zero-field splitting is observed experimentally. We attribute this splitting to some anisotropy in the x-y plane breaking this symmetry and allowing a preferred magnetization axis of the YIG at 0 field. Some possible explanations for this anisotropy are a small visible damage to our YIG crystal on one edge or possible imperfections in eccentricity and alignment. If the magnetic domains of unsaturated YIG preferentially align with one in-plane axis 
compared to its orthogonal axis within the $x$-$y$ plane, this anisotropy would result in a relative frequency shift between the standing-wave modes along the in-plane easy and hard axes.  This anisotropy-induced frequency shift $\pm\beta$ for the $x$ and $y$ modes can be modeled in numerical simulations employing a permeability 
tensor of unsaturated ferromagnets~\cite{schlomann_microwave_1970, green_microwave_1974} with certain anisotropic assumption, which can plausibly explain the data (Appendix A). As $B$ increases, we expect $\beta$ to decay towards 0 when the magnetic domains are increasingly aligned towards the $z$ direction, thus making any $x$-$y$ plane energetic preference of negligible effect.  We model this decay with a thermodynamic toy model (Appendix B) whose details do not affect the conclusions of this study.

For the rest of this article, we will focus on the pair of polariton modes near 11 GHz in Fig.~\ref{fig:internal_spectrum}(b), and refer to them as 
``the circulator modes" for reasons that will become apparent. 
We can model their frequencies in the partially magnetized regime ($|B|< 50$ mT) using a phenomenological model accounting for the degeneracy-lifting anisotropy and the field-dependent magnetization of YIG. Let the zero-field frequencies of the $x$ and $y$ modes be $\omega_x=\omega_y$ if the device had perfect 3-fold symmetry,  $\beta$ and $\theta/2$ be the magnitude of anisotropy caused degeneracy-lifting and the direction of the in-plane anisotropy axis (relative to the $x$ axis), and off-diagonal imaginary coupling term  $\pm ikB$ be the magnetic field induced degeneracy-lifting, linearly increasing with a real coefficient $k$. We use the following Hamiltonian to characterize the pair of circulator modes in the basis of $x$ and $y$ mode amplitudes:
\begin{equation}
H/\hbar=
\begin{pmatrix}
\omega_{x} + \beta\cos{\theta} + mB^{2} & \beta\sin{\theta} + ikB \\
\beta\sin{\theta} – ikB & 
\omega_{y} - \beta\cos{\theta} + mB^{2} 
\end{pmatrix}
\end{equation}
This effective model of the polariton modes has absorbed the magnon contributions in the regime where they have been adiabatically eliminated. The formation of clockwise and counterclockwise eigenmodes is due to magnon-mediated interactions modeled by $\pm ikB$. 
The level repulsion from the far-detuned magnon modes is approximated by a small quadratic shift in frequency $mB^{2}$.
The quadratic dependence was empirically chosen because the sum of the mode frequencies over field displayed a roughly quadratic relationship with $B$ over the plotted field range. By fitting the mode spectrum in Fig.~\ref{fig:internal_spectrum}(b), we obtain $\omega_x/2\pi=\omega_y/2\pi=11.054$ GHz, $k/2\pi=9.82$ GHz/T, $m/2\pi=50$ GHz/T$^2$, $\beta/2\pi=139$ MHz.  

\begin{figure*}[tbp]
    \centering
    \includegraphics[scale=.41]{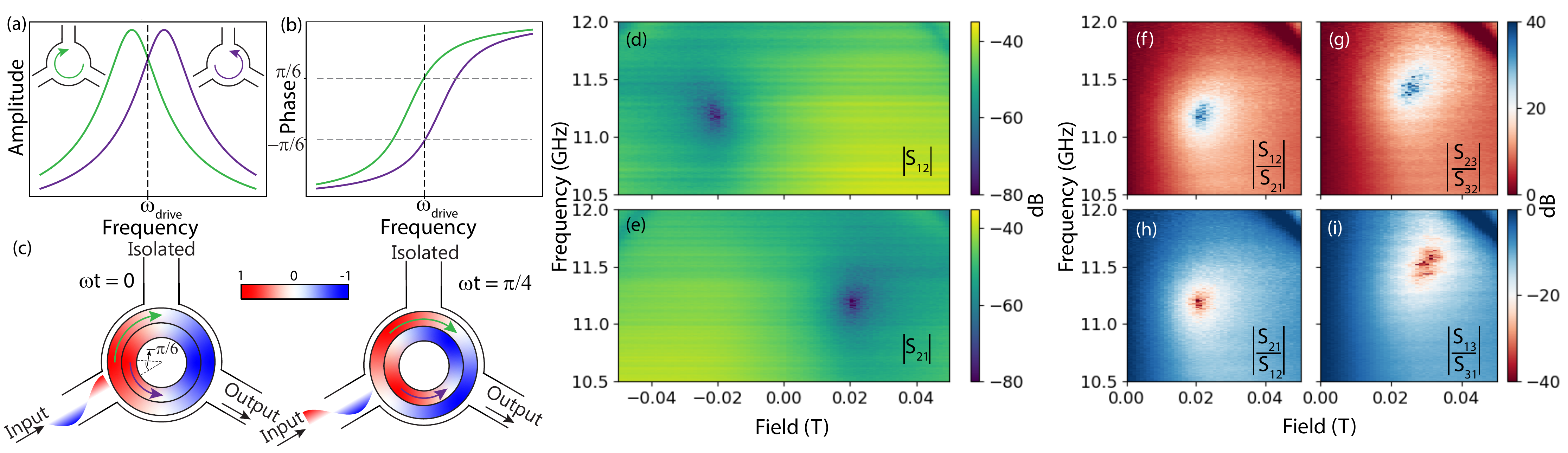}
    \caption{\textbf{Illustration of the circulator working principle and low-temperature characterization of the non-reciprocity.} In the circulator package with IMT, the frequency splitting of clockwise and counterclockwise rotating modes as shown in (a) can be tuned such that the phase of the modes are $\pi/6$ and $-\pi/6$ as shown in (b). This then produces a node at the upper port, thereby preventing any signal from leaving there at all times where $\omega t=0$ and $\omega t=\pi/4$ are shown pictorially in (c).  (d, e) Measured microwave transmission (d) $|S_{12}|$ and (e) $|S_{21}|$ spectra as a function of magnetic field B. (f-i) The isolation performance, (f) $\mathcal{I}_{12}=|S_{12}/S_{21}|$,(h) $\mathcal{I}_{21}=|S_{21}/S_{12}|$, (g) $\mathcal{I}_{23}=|S_{23}/S_{32}|$, (i) $\mathcal{I}_{13}=|S_{13}/S_{31}|$. $S_{21}$ is obtained by measuring the $S_{12}$ at –$B$, which provides a self-calibrated way to determine the isolation of the circulator.
    \label{fig:circulator}
    }
\end{figure*}

It is well-known that the FMR modes of partially magnetized ferrimagnetic insulators, where the magnetic domains are not aligned in equilibrium, have large damping.  Therefore, one may expect broad linewidths for photon-magnon polariton modes below magnetic saturation.  Indeed, we have observed linewidths exceeding 100 MHz for another polariton mode at 5 GHz at $B<$ 50 mT (not shown). Surprisingly, the polariton modes at higher frequency display narrow linewidths, $\kappa_i\approx2$ MHz for the pair of circulator modes [Fig.~\ref{fig:internal_spectrum}(c)], which corresponds to quality factors on par with some circuit QED elements such as the readout resonators.  
The narrow internal linewidth of the circulator modes is crucial for constructing a low-loss circulator and eventually achieving high quantum efficiency of non-reciprocal interactions in circuit QED.  It is primarily aided by the use of single crystalline YIG and the relatively low magnon participation in the circulator modes compared to commercial circulators.  The observed $\kappa_i$ may be limited by either the spin relaxation in YIG or the Ohmic loss in copper.  The former remains to be investigated in this partially magnetized regime, and the latter may be further reduced through better surface treatment or the use of superconducting materials in low-field regions of the waveguide package.

\section{Circulator characterization}

The device acts as a circulator when the end of each waveguide section is in IMT rather than WCP with an applied magnetic field in the $\hat{z}$ direction. In this configuration, the linewidths of all internal modes are substantially broadened forming a transmission continuum in the measurement, as shown in Fig.~\ref{fig:circulator}(d,e) for $|S_{12}|$ and $|S_{21}|$.  Nevertheless, the operating condition of the circulator can be conceptually understood as having a pair of counter-propagating internal modes with their magnetic-field-induced splitting ($\delta$) satisfying the relationship $\delta = 2\kappa_c/{\sqrt{3}}$ versus their half linewidths ($\kappa_c$)~\cite{fay_operation_1965}. As illustrated in Fig.~\ref{fig:circulator}(a-c), when driven at a frequency in the middle of the two resonances, the two circulator modes are excited with equal amplitude and a phase shift of $\pm\frac{\pi}{6}$ relative to the drive.  The resultant standing wave pattern forms a node at the isolation port of the circulator.  This condition can be satisfied by choosing the correct combination of frequency and magnetic field.

We characterize the non-reciprocity of the circulator by the isolation ratio $\mathcal{I}_{12}=|S_{12}/S_{21}|$, which may be computed from Fig.~\ref{fig:circulator}(d,e). However, since $S_{12}$ and $S_{21}$ are measured through different cables and amplifier chains [Fig.~1(b)], it is challenging to calibrate their absolute values precisely.  A much better self-calibrated technique to extract the isolation ratio in our system is to use the Onsager-Casimir relation~\cite{casimir_onsagers_1945}, $S_{21}(B)=S_{12}(-B)$, resulting from the microscopic time reversal symmetry.  
Therefore, we use $\mathcal{I}_{12}=|S_{12}(B)/S_{12}(-B)|$ to determine the isolation ratio of the circulator as shown in Fig.~\ref{fig:circulator}(f), with the (field-independent) contribution from same transmission chain cancelled out.  The result indicates the circulator working condition is met for the pair of counter-propagating modes at $\sim$11.2 GHz with external field $\sim$0.022 T.  We see $\geq$20 dB of isolation over a bandwidth of about 250 MHz, with maximum isolation of at least 35 dB.  
 
The same analysis on $S_{21}$ data yields the same isolation property [Fig.~\ref{fig:circulator}(h)] as expected. Similarly, $\mathcal{I}_{23}$ and $\mathcal{I}_{13}$ are measured as in Fig.~\ref{fig:circulator}(g) and (i), each showing a slightly different working field and frequency (possibly due to imperfections of the device geometry) but similar isolation magnitude and bandwidth.  These data are measured at an estimated circulating photon number on the order of 10's, but when we lower the power to below single photon level, the isolation property does not show notable changes.

An important motivation of our work is to ultimately implement pristine non-reciprocal interactions between superconducting qubits or cavities. It is crucial to minimize the ratio between the undesirable internal dissipation ($\kappa_i$) and the external bath coupling ($\kappa_c$) that enables non-reciprocity.  In the case of a circulator, this ratio sets the limit for the circulator's microwave insertion loss $\mathcal{L}_{21}$~\cite{kord_microwave_2020, fay_operation_1965}:
\begin{equation}
    \mathcal{L}_{21}=1-|S_{21}|^2 \geq 1-|S_{21}|^2-|S_{11}|^2-|S_{31}|^2 \approx \frac{\kappa_i}{\kappa_c}
\end{equation}

\begin{figure}[tbp]
    \centering
    \includegraphics[scale=.42]{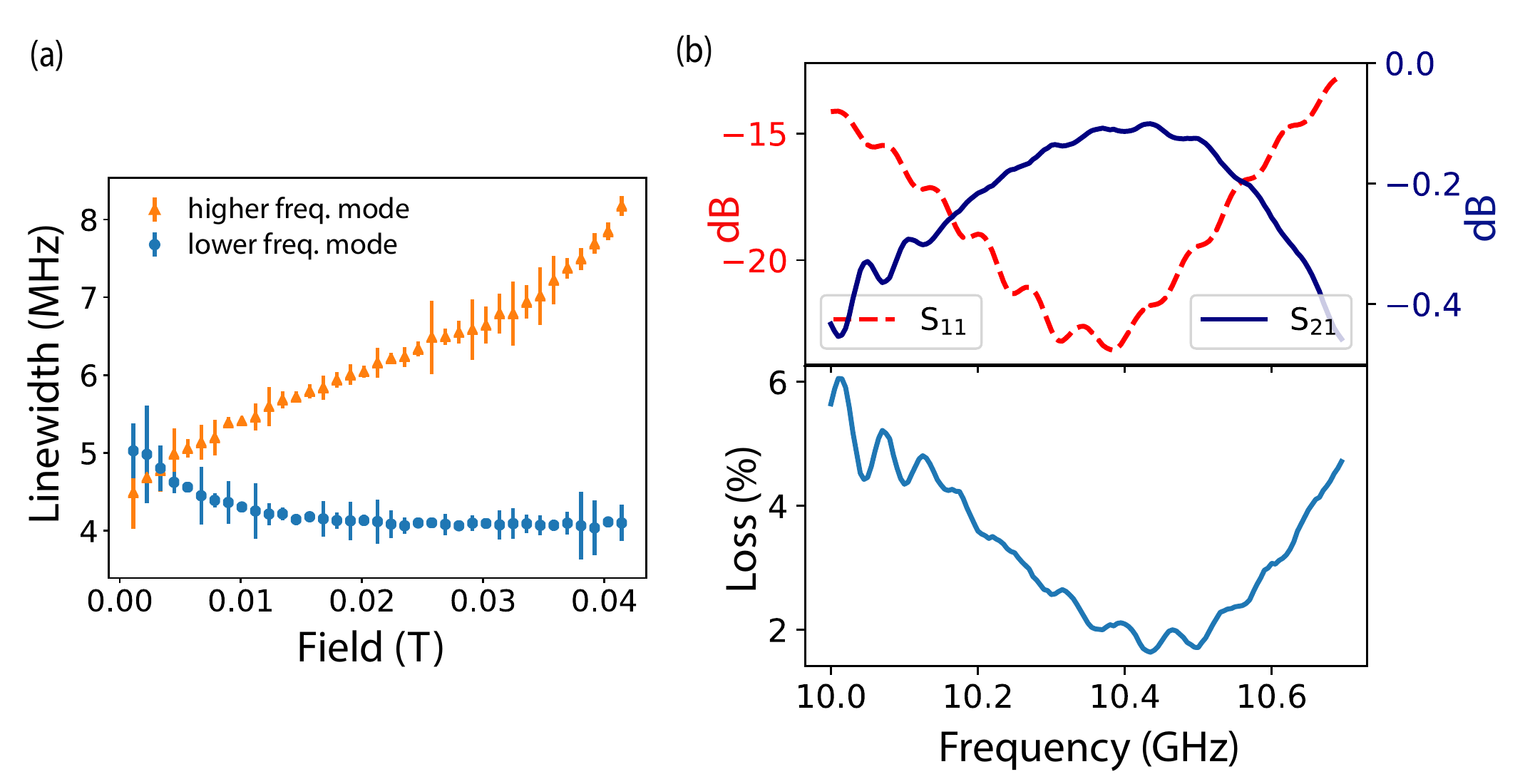}
    \caption{\textbf{Characterization of the internal loss of the circulator at room temperature.} (a) Linewidths of the pair of circulator modes measured at room temperature.  (b) Transmission $S_{21}$ and reflection $S_{11}$ near the maximum isolation regime of the circulator, measured at $B = 24.8$ mT (top panel) and the internal loss of the circulator calculated from it (bottom panel) at room temperature.
    \label{fig:RT_loss}
    }
\end{figure}

This lower limit is obtained in principle when the circulator has perfect impedance matching ($S_{11}=0$) and isolation ratio ($S_{31}=0$).  Typical commercial ferrite circulators used in circuit QED experiments have shown insertion loss around 10\%~\cite{kurpiers_deterministic_2018, axline_-demand_2018}, which is dominated by internal loss. Experimental Josephson circulators so far have also reported insertion loss of -0.5 dB (11\%) or higher~\cite{chapman_widely_2017, lecocq_nonreciprocal_2017}.  The lowest quoted insertion loss for a commercially-listed waveguide circulator is -0.1 dB (or 2.2\%) but that is untested in the quantum regime.   
In order for the quantum efficiency of a non-reciprocal two-qubit interaction channel to match the fidelity of state-of-the-art two-qubit operations, the insertion loss would need to be improved to the sub-percent level.

To the best of our knowledge, it is an open challenge to calibrate the insertion loss of a microwave component in a dilution refrigerator with a precision better than 1\%. Even using specialized Thru-Reflect-Line calibration components and well-characterized cryogenic switches, the resultant precision would still be limited to about 0.1dB (or 2.3\%)~\cite{ranzani_two-port_2013}. In order to infer the loss of our circulator at 20 mK, we measure its $S$ parameters at room temperature after a careful calibration procedure that uses attenuators in series to suppress standing waves.  We find a conservative upper bound for room-temperature internal loss of $\leq 1 – |S_{21}|^2 – |S_{11}|^2  \approx 2\%$, as shown in Fig.~\ref{fig:RT_loss}(b).  Assuming $\kappa_c$ does not change as a function of temperature, comparing the intrinsic linewidth of the circulator mode pair at room temperature versus 20 mK would inform the internal loss at 20 mK.  The intrinsic linewidths, measured 
in WCP, are 4.1 and 6.3 MHz at room temperature [Fig.~\ref{fig:RT_loss}(a)] and 1.8 and 2.2 MHz at low temperature [Fig.~\ref{fig:internal_spectrum}(c)], indicating that $\kappa_c\gtrsim 260$ MHz and $\kappa_i/\kappa_c\lesssim 0.8\%$.  If we instead use the relation of $\kappa_c=\sqrt{3}\delta/2$, which yields $\kappa_c$ in the range of 430 MHz to 550 MHz (and data in Section V would further suggest $\kappa_c$ at the high end of this range), or  $\kappa_i/\kappa_c\approx0.4\%$. 
Further improvement of the circulator bandwidth and the coupling ratio can be achieved by applying impedance transformation techniques to increase $\kappa_c$~\cite{helszajn_stripline_2008}.  

Translating this small internal loss ratio to a sub-percent insertion loss for a circulator as a peripheral transmission-line device would further require excellent impedance matching.  However, we emphasize that this requirement is not fundamental if the circulator is modeled as part of the quantum system itself mediating interactions between other quantum resonance modes. Unlike most ferrite circulators, our device operates in the regime of partial magnetization for YIG.  It only requires a moderate external magnetic field that is significantly below the critical field of a variety of superconducting materials.  This allows for 3D integration of superconducting niobium cavities and shielded transmon qubits for studying circuit QED with non-reciprocal interactions. In the following section, we demonstrate direct coupling of two external superconducting cavity modes with the circulator modes and analyze the resultant non-reciprocal hybrid system as a whole.

\section{Tuning non-reciprocity of eigenmode structure}

\begin{figure*}[tbp]
    \centering
    \includegraphics[scale=.34]{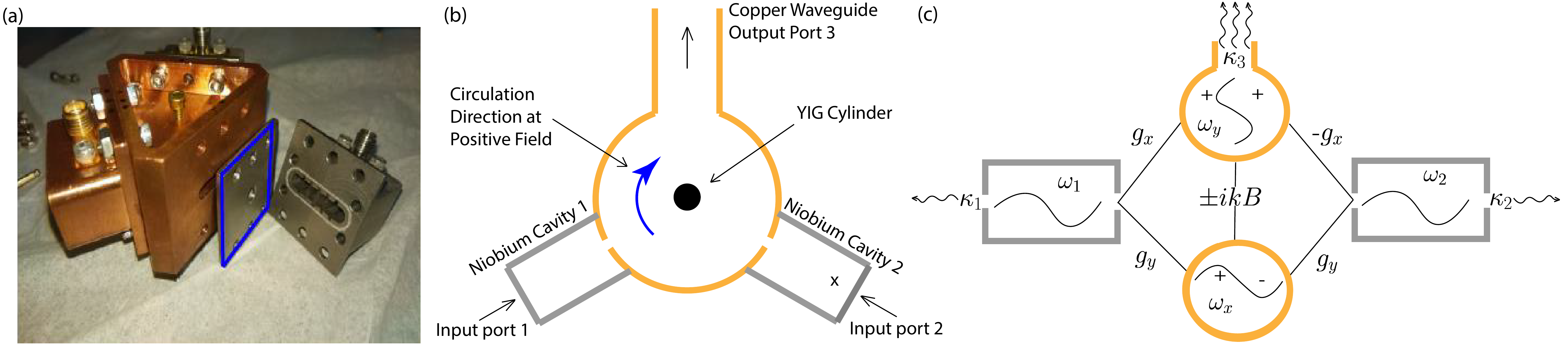}
    \caption{\textbf{Waveguide circulator-cavity integration.}  (a) The photo image, (b) a schematic top-down view, and (c) a diagrammatic illustration of the effective Hamiltonian (see Eq.~(4), for clarity the $\beta$ and $mB^2$ terms have been neglected in the illustration) of our integrated non-reciprocal device.  It is composed of a Cu waveguide Y-junction loaded with a YIG cylinder, two Nb cavity segments with weakly-coupled drive ports (Port 1 and 2), and an output port with IMT (Port 3). For each cavity, the sidewall closest to the copper Y-junction is formed by a standalone niobium plate in the assembly [enclosed in blue in (a)], which contains a 5 mm-diameter aperture to create an evanescent coupling between the superconducting cavity mode and the circulator modes.  One of the cavities is loaded with a transmon qubit [marked as $\times$ in (b)] which stays unused in its ground state in this study. 
    \label{fig:cavity_integration}
    }
\end{figure*}

We integrate superconducting cavities with the ferrite device by replacing the rectangular waveguide extensions with superconducting 3D cavities made of niobium [Fig.~5(a)]. 
Two cavities, attached at Port 1 and 2, are tuned to have resonance frequencies close to each other, $\omega_1\approx\omega_2\sim 10.8$ GHz, both of which are within the bandwidth of the circulator.  Each cavity is coupled to the central Y-junction via a coupling aperture.  As a result, the circulator modes will mediate an interaction between these two external cavities. Crucially, this circulator-mediated interaction can have both coherent and dissipative aspects, and can be non-reciprocal.  The degree and the direction of non-reciprocity of the coupling can be tuned via the external magnetic field.  Note that Port 3 remains impedance-matched to a transmission line. This is also essential:  it serves as the dominant dissipative bath that is necessary for achieving non-reciprocal inter-mode interactions~\cite{metelmann_nonreciprocal_2015}. 

\begin{figure*}[tbp]
    \centering
    \includegraphics[scale=.32]{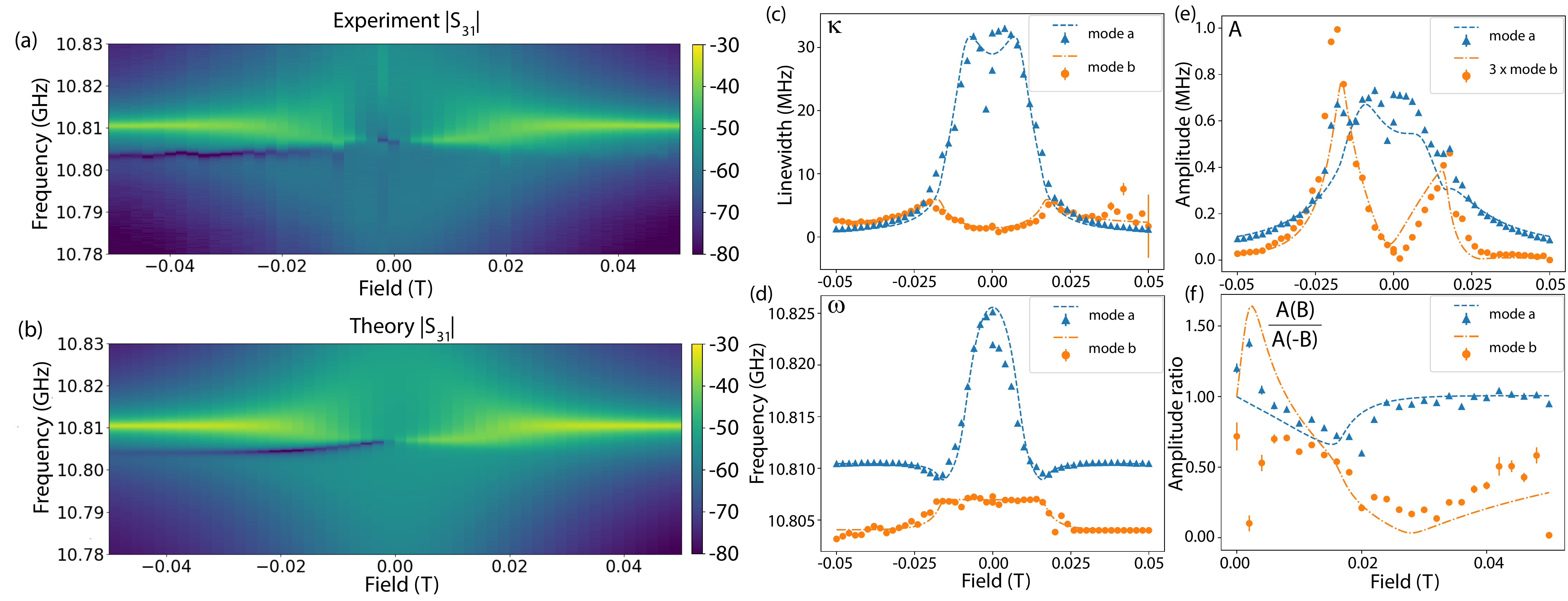}
    \caption{\textbf{Spectroscopy of the hybridized non-reciprocal modes of a circulator-cavity system.} (a) VNA transmission measurement and (b) model prediction of $|S_{31}|$ frequency spectrum over external magnetic field $B$.  Remaining panels show magnetic field dependence of the system's eigenmodes and wavefunctions:   (c) eigenmode linewidths $\kappa_n/2\pi$, (d) eigenmode frequencies $\omega_n/2\pi$, (e) amplitude parameter $A_n$ (c.f.~Eq.(3)), and (f) amplitude ratio (c.f.~Eq.(8)) of experimental data (dots) from two-mode Lorentzian fit (Eq.(3)) and theory predictions (dashed lines). The symmetry of $\kappa_n$ and $\omega_n$ ($i.e.~$ complex eigen-energy of the hybrid system) with respect to $B$ exemplifies the microscopic time-reversal symmetry of the non-Hermitian system.  The non-reciprocity is reflected in the difference in $A_n$ at $\pm B$, which reveals the asymmetry in the left/right eigen-vector structure (c.f.~Eq.(8))).
    The effective Hamiltonian parameters from the fit are: $\omega_1/2\pi=10.8104$ GHz, $\omega_2/2\pi=10.8040$ GHz, $\omega_x/2\pi=10.707$ GHz, $\omega_y/2\pi=10.813$ GHz, $\theta=37.7^{\circ}$, $\kappa_3/2\pi=730$ MHz, $g_x/2\pi=(9.0 +0.011\beta)$ MHz, $g_y/2\pi=(5.0+0.006\beta)$ MHz with $\beta/2\pi=139$ MHz at $B=0$ and decays with $|B|$.
    \label{fig:hybrid_spectrum}
    }
\end{figure*}

To probe the hybridized mode structure of the composite system, we measure $S_{31}$ from a weakly-coupled drive port on Cavity 1 to Port 3. The measured amplitude of $S_{31}$ as a function of magnetic field and frequency is shown in Fig.~\ref{fig:hybrid_spectrum}(a).  There are a total of four bare oscillator modes in the vicinity (within 0.5 GHz) of the frequency range of interest: two superconducting cavity modes and two internal circulator modes.  Since the loaded circulator modes with very broad linewidths ($>100$ MHz) are difficult to observe in the presence of the standing-wave background of the coaxial cables, this spectroscopy measurement primarily reveals the eigenmodes that are localized in the external superconducting cavities.  Indeed, at $|B|>0.03$ T, the spectrum shows two sharp resonances which we identify as the bare cavity modes to a good approximation. At lower fields, the cavities appear to more strongly hybridize with each other and with the lossy circulator modes, but the spectrum can nontheless be captured relatively well by the sum of two Lorentzian modes $a$ and $b$: 
\begin{equation}
    S_{31} = \frac{A_{a}e^{i\phi_{a}}}{{-i(\omega-\omega_{a})-\kappa_{a}/2}}+\frac{A_{b}e^{i\phi_b}}{{-i(\omega-\omega_{b})-\kappa_{b}/2}}
    \label{eq:Lorentzian}
\end{equation}
By fitting the spectrum to Eq.~(\ref{eq:Lorentzian}), we can extract the  linewidth ($\kappa_{i}$), frequency ($\omega_{i}$) and amplitude ($A_{i}$) of the two Lorentzians at each magnetic field, as plotted in Fig.~\ref{fig:hybrid_spectrum}(c-e).\\

The magnetic field dependence of the two prominent Lorentzians can be connected to the eigenmode solutions of an effective Hamiltonian model of system.  We describe the system using the following 4$\times$4 non-Hermitian matrix, written in the basis of the amplitudes of the two cavity modes and the two circulator modes: 
\begin{widetext}
\begin{equation}
H_{\mathrm{eff}}/\hbar=\\
\begin{pmatrix}
\;\omega_1-i\frac{\kappa_1}{2}\; & 0 & g_y & g_x\\
0 & \;\omega_2-i\frac{\kappa_2}{2}\; & g_y & -g_x\\
g_y & g_y & \omega_y-\beta\cos\theta+mB^{2}-i\frac{\kappa_3}{2} & \beta\sin\theta - ikB\\
g_x\ & -g_x\ &  \beta\sin\theta + ikB & \omega_x+\beta\cos\theta+mB^{2}\\
\end{pmatrix}
\label{eq:4x4}
\end{equation}
\end{widetext}
The two niobium cavities have bare frequencies $\omega_1$, $\omega_2$, and input coupling rates of $\kappa_1$ and $\kappa_2$.  The bottom right block of Eq.~(4) describes the two circulator modes, with their anisotropy dependence and imaginary coupling due to magnon hybridization following the same description as in Eq.~(1).  
The zero-field frequencies of the two circulator modes $\omega_x$, $\omega_y$ are no longer equal since the device is no longer 3-fold symmetric.  The $y$-mode with frequency $\omega_y$ is symmetric with respect to the $y$ axis, and therefore has an equal and in-phase coupling rate $g_y$ with the two cavities.  It rapidly leaks to the waveguide output Port 3, with a decay rate $\kappa_3\gg\kappa_1, \kappa_2, g_x, g_y$.  $\kappa_3$ is related to $\kappa_c$ of the loaded circulator as in Section IV by $\kappa_3=4\kappa_c/3$. The $x$-mode is anti-symmetric with respect to the $y$ axis, preventing it from coupling to the output port. This also leads to a 180$^{\circ}$ phase difference in cavity coupling as accounted for by the negative sign on two of the $g_x$ parameters

This effective non-Hermitian Hamiltonian can be diagonalized as:
\begin{equation}
H_\mathrm{eff} = \sum_{n}\hbar\omega_{n}\ket{n_{R}}\bra{n_{L}}
\end{equation}
where $n=a,b,c,d$ are the eigenmode indices of the system, $\omega_n$ the complex eigen-frequencies, and $\ket{n_R}$ and $\ket{n_L}$ the right and left eigenvectors of the non-Hermitian Hamiltonian, defined as: $H_\mathrm{eff}\ket{n_{R}}=\hbar\omega_{n}\ket{n_{R}}$ and $H_\mathrm{eff}^{\dagger}\ket{n_{L}}=\hbar\omega_{n}^*\ket{n_{L}}
$.  The scattering matrix element $S_{ij}$ from Port $j$ to Port $i$ can be generally drived from the input-output theory relation:
$S_{ij}=\delta_{ij}-i\sqrt{\kappa_{i}\kappa_{j}}G_{ij}^{R}(\omega)$,
where the $4 \times 4$ retarded matrix Green's function is defined as: $G^{R}(\omega)=(\omega-H_\mathrm{eff})^{-1}$, and $\kappa_i$ and $\kappa_j$ are the output and input coupling rates, respectively.  Applying this formalism to the $S_{31}$ measurement of our device, we arrive at the following Lorentzian spectral decomposition to describe the spectrum: 
\begin{equation}
S_{31}(\omega)=\sum_{n}\frac{-i\sqrt{\kappa_{1}\kappa_3}\bra{y}\ket{n_{R}}\bra{n_{L}}\ket{1}}{\omega-\omega_n}
\end{equation}
where the real and imaginary parts of the eigen-frequency $\omega_n$ correspond to the observed Lorentzian frequencies and half linewidths, respectively.  The amplitudes of the Lorentzians are proportional to the product of the left eigenvector overlap with the bare cavity mode $\ket{1}$ and the right eigenvector overlap with the output circulator mode $\ket{y}$.

By fitting the extracted Lorentzian parameters of the two prominent eigenmodes in Fig.~\ref{fig:hybrid_spectrum}(c-e) to the predictions of the $4\times4$ Hamiltonian model across all fields (Eq.~4), we can determine all the free Hamiltonian parameters in this model.  This includes $\kappa_3=730$ MHz, implying $\kappa_c=550$ MHz for the loaded circulator, consistent with (and at the high end of) the estimates in Section IV.  Somewhat surprisingly, the experimental data strongly suggests that the cavity-circulator coupling rates $g_x$ and $g_y$ must be magnetic field dependent. (For example, it heavily constrains that $g_x/2\pi > 16$ MHz near $B$ = 0 and $g_x/2\pi<12$ MHz at $|B|>30$ mT.)  We attribute this varying coupling to the change in electromagnetic field distribution of the $x$- and $y$-modes around the coupling aperture due to the $x$-$y$ anisotropy of YIG.   Assuming $g_x$ and $g_y$ contains a contribution proportional to $\beta(B)$ with the same decay shape over applied field, the effective Hamiltonian model fits the Lorentzian parameters quite well and also reproduces the overall transmission spectrum [Fig.~\ref{fig:hybrid_spectrum}(b)]. 

The eigenmode features of the system can be understood intuitively by considering first the inter-mixing of the $x$, $y$ circulator modes (i.e.~diagonalization of the lower right block of $H_{\mathrm{eff}}$) and then their mixing with the two cavity modes.  At $B = 0$, the circulator modes are relatively close in frequency to the bare cavities, resulting in strong four-mode hybridization and substantial linewidth-broadening and frequency shift to Mode $a$.  
As $B$ increases, the block-diagonalized circulator modes split further in frequency in response to increasing magnetization of YIG (analogous to the unloaded internal mode spectrum in Fig.~2b), and become more detuned from the bare cavities, so the cavity-circulator hybridization is continuously reduced. This is reflected in the eventual flattening of the frequency and linewidth of the observed Lorentzians at high fields.  

In our device, opposite magnetic fields produce opposite directions of non-reciprocity, hence the transmission spectra observed at $\pm B$ in Fig.~\ref{fig:hybrid_spectrum}(a) are markedly different.  Interestingly, the extracted data in Fig.~\ref{fig:hybrid_spectrum}(c,d) shows that the underlying eigenmode frequency and linewidths at $\pm B$ are equal, unchanged under the mapping $\mathcal{P}$ of $B : \mapsto -B$.  This is no coincidence, but is rather the direct consequence of microscopic symmetry requirements.  Recall again that the Onsager-Casimir relation~\cite{casimir_onsagers_1945} requires that the full 
scattering matrix $S$ satisfy $S(-B) = S^\mathrm{T}(B)$.  As $S$ is however directly determined by our non-Hermitian Hamiltonian, this necessarily implies that $H^{\phantom{T}}_\mathrm{eff}(-B)=H_\mathrm{eff}^\mathrm{T}(B)$.  This in turn implies that the complex eigenvalues of $H_\mathrm{eff}$ are unchanged under $\mathcal{P}$.  Note that the operation $\mathcal{P}$ is not just a simple time-reversal operation, as it does not involve transforming loss to gain (and vice versa).       
This property of $H_\mathrm{eff}$ can be easily seen to hold for our specific model in Eq.~(4).  Nonetheless, we emphasize that our experimental observation here of eigenvalue invariance under the mapping $\mathcal{P}$ is a demonstration of a general physical property; it is by no means contingent on the specifics of our model.

While the eigenvalues of $H_\mathrm{eff}$ do not directly reflect the non-reciprocal physics of our system, the same is not true of its eigenvectors. As it involves matrix transposition, the operation $\mathcal{P}$ exchanges the left and right eigenvectors of the effective Hamiltonian: $\ket{n_R( B)}=\ket{n_L(-B)}^{*}$.  A defining feature of a non-reciprocal Hamiltonian is that the left and right eigenvectors generally differ in their spatial structures
(i.e.~they look very different when expressed in a basis of bare modes):
\begin{equation}
R_{i,n} = \frac{|\bra{n_L}\ket{i}|}{|\bra i \ket{n_R}|}\neq 1
\end{equation}
As has been discussed elsewhere~\cite{mcdonald_exponentially-enhanced_2020, schomerus_nonreciprocal_2020}, the $R_{i,n}$ characterizes a fundamental asymmetry in the response of our system.  The numerator characterizes the susceptibility of the eigenmode $n$ to a perturbation or excitation entering from bare mode $i$.  In contrast, the denominator tells us the amplitude on bare mode $i$ that would result given that the system eigenmode $n$ is excited.  In a Hermitian system these quantities are necessarily identical, expressing a fundamental kind of reciprocity between susceptibility and response.  In our non-Hermitian system, the non-unity ratio here reflects the effective non-reciprocity of the inter-mode interactions.

This non-reciprocal eigenvector structure is experimentally verified by the asymmetry of the Lorentzian amplitudes with respect to $B$ in Fig.~\ref{fig:hybrid_spectrum}(e), 
\begin{equation}
\frac{A_n(B)}{A_n(-B)}=\bigg|\frac{\bra{n_L}\ket{1}}{\bra{1}\ket{n_R}}\frac{\bra{y}\ket{n_R}}{\bra{n_L}\ket{y}}\bigg| = \frac{R_{1,n}}{R_{y,n}}
\label{eq:ampratio}
\end{equation}
We plot this ratio in Fig.~\ref{fig:hybrid_spectrum}(f).  For our device, a calculation based on Eq.~(\ref{eq:4x4}) shows that $R_{y,n}\approx 1$ for most of the field range (near zero field and $|B| \gtrsim 15$ mT), allowing Fig.~\ref{fig:hybrid_spectrum}(f) to be understood as a measurement of the non-reciprocity ratio $R_{1,n}$, in this field range, showing the role of Cavity 1 in the two prominent eigenmodes of the system. 
In particular, the most pronounced asymmetry is observed near the optimal working point of the circulator ($B=\pm28$ mT) for Mode $b$, which can leak through cavity 1 but cannot be excited from Cavity 1 or vice versa, as expected for a mode dominated by photons in Cavity 2.

\begin{figure}[tbp]
    \centering
    \includegraphics[scale=.4]{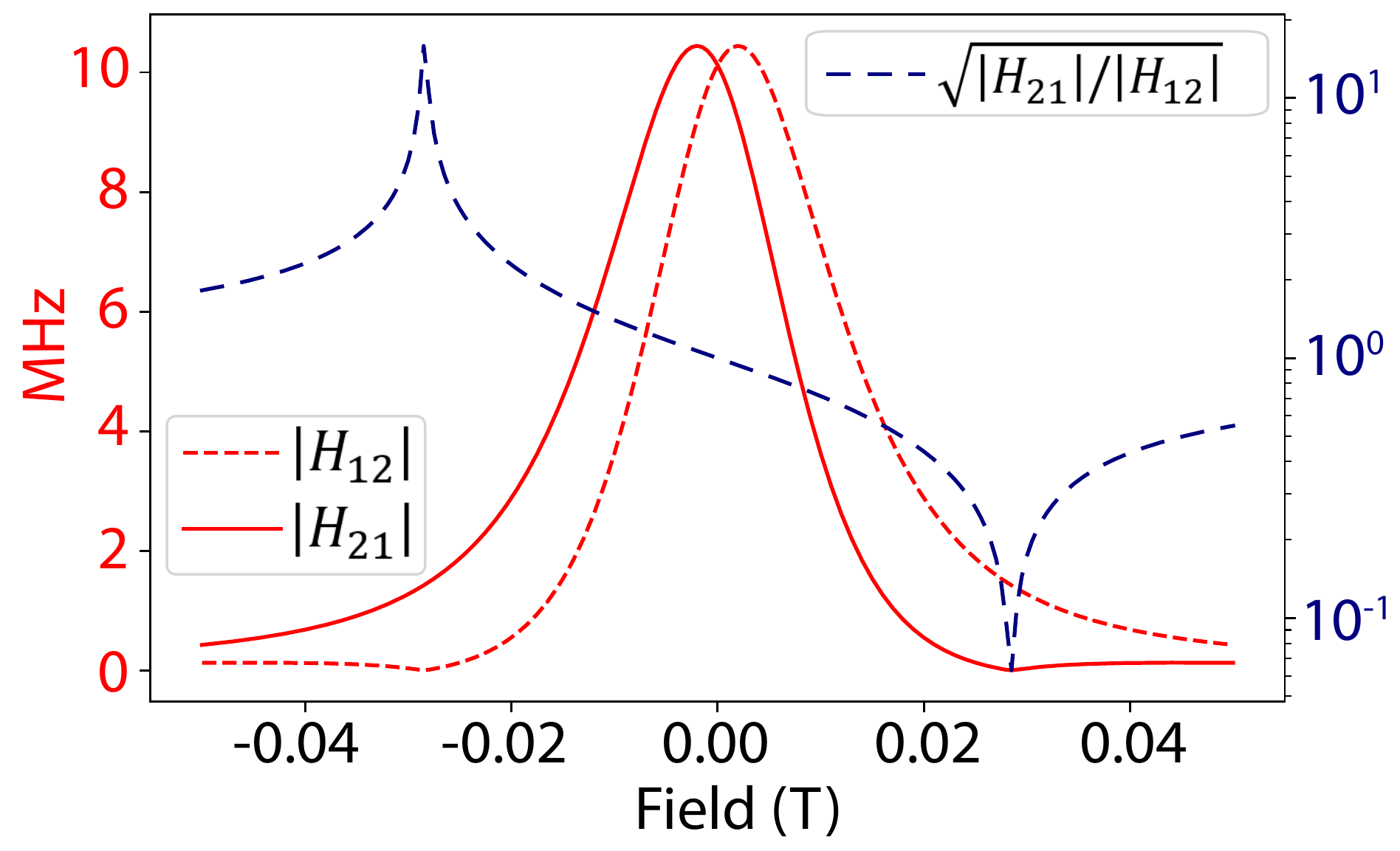}
    \caption{\textbf{Mediated non-reciprocal coupling rates between the external superconducting cavities.} Red curves show the off-diagonal coupling terms in the effective two-mode Hamiltonian (c.f.~Eq.(9)), $\lvert H_{12}\rvert$ and $\lvert H_{21}\rvert$, as a function of magnetic field, and blue shows $r=\sqrt{\lvert H_{21}\rvert/\lvert H_{12}\rvert}$.
    \label{fig:2by2 matrix}
    }
\end{figure}

It is also interesting to discuss our system in the context of general systems exhibiting non-reciprocal interactions between constituent parts.  Such systems are commonly described by phenomenological non-Hermitian Hamiltonian matrices $H$, whose matrix elements in a local basis encode interactions with directionality: $|H_{ij}| \neq |H_{ji}|$.  A prominent example is the Hatano-Nelson model~\cite{hatano_vortex_1997} of asymmetric tunneling on a lattice. In our case, we have a microscopically-motivated model that is fully consistent with the requirements of microscopic reversability, but which encodes non-reciprocity.  As shown in appendix C, one can adiabatically eliminate the internal circulator modes from our system to obtain an effective two-mode non-Hermitian Hamiltonian that describes the external cavity modes and their circulator-mediated interaction:
\begin{align}
H'_{\mathrm{eff}}/\hbar=
\begin{pmatrix}
\omega_{1,\mathrm{eff}} -i\frac{\kappa_{1,\mathrm{eff}}}{2} & H_{12}\\
H_{21} & \omega_{2,\mathrm{eff}} -i\frac{\kappa_{2,\mathrm{eff}}}{2}\\
\end{pmatrix}
\end{align}
The field-tunable non-reciprocity can be seen in the asymmetry of the off-diagonal coupling values $H_{12}$ and $H_{21}$ based on the model, as plotted in Fig.~\ref{fig:2by2 matrix}. We note that the scale of $H_{12}$ and $H_{21}$ of a few MHz (which can be increased by using a larger coupling hole) is much larger than the achievable internal loss of the superconducting cavities , making the non-reciprocal coupling the dominant interaction. 

Furthermore, this reduced Hamiltonian and its eigenvectors can be mapped to a reciprocal Hamiltonian and associated eigenvectors using a similarity transformation $S(r)$, where $\sqrt{\lvert H_{21}\rvert/\lvert H_{12}\rvert} \equiv r$, as outlined in Appendix C. The similarity transform effectively localizes the mode participation on the lattice site in the direction of stronger coupling more than would be expected in the reciprocal case, causing the amplitude ratios ($R_{i,n}$) to deviate from 1, which can be viewed as a consequence of the non-Hermitian skin effect on a two site Hatano-Nelson lattice~\cite{yao_edge_2018}. Furthermore,the similarity transformation can be used to explain the qualitative behavior of the disparate amplitude ratios seen in Fig.~\ref{fig:hybrid_spectrum}(f). In particular, for $B\gtrsim20$ mT, we have $R_{1,a}\approx 1$ and $R_{1,b}\approx r^2$ (see Appendix C for details).

\section{Outlook}

In this work, we have revisited the working principles of a Y-junction ferrite circulator~\cite{fay_operation_1965}, a microwave engineering classic from the 1960’s, in an entirely new context of hybrid quantum systems and non-Hermitian Hamiltonian.  The use of reconfigurable probes and single-crystalline YIG in a low-loss waveguide package allows us to connect the properties of the photon-magnon polaritons to the circulator performance.  We have further leveraged our direct access to the internal modes and our ability to tune their coupling \textit{in situ} to construct a multi-mode chiral system and unambiguously reveal its non-reciprocal eigenvector structure. An understanding of the circulator modes and the non-reciprocal eigen-vector structure of the multi-mode chiral system provides a foundation for future engineering of any target non-Hermitian Hamiltonian. This is achieved by our creation of a template model that one can use to couple any circuit QED element to in order to understand how it would integrate into the non-reciprocal dynamics, as was done here with the superconducting cavities.

Looking forward, our device architecture provides a versatile testbed for studying non-reciprocal interactions in circuit QED by integration of superconducting qubits.  This is enabled by two of its highlighted properties: the low internal loss of the circulator modes ($<$1\% of the demonstrated coupling rates, compatible with potential high-fidelity operations), and the relatively low-field operation of the circulator ($\sim$25 mT, below ferrimagnetic saturation).  The latter allows niobium waveguides or cavities to conveniently act as magnetic shields for superconducting qubits.  We have preliminarily tested that the coherence times of a transmon qubit housed in one of the niobium cavities are unaffected by in-situ application of a global magnetic field up to at least 0.1 T.  We expect a transmon housed in a niobium waveguide should receive a similar level of protection from magnetic field. 

Direct non-reciprocal coupling of superconducting qubits would open a new frontier in the study of non-reciprocal dynamics currently dominated by linear systems~\cite{sounas_non-reciprocal_2017,fang_generalized_2017, xu_nonreciprocal_2019, ruesink_nonreciprocity_2016}. 
The physics of a $N$-mode linear non-reciprocal system can always be described efficiently by a $N\times N$ non-Hermitian Hamiltonian matrix (exemplified by our application of such a model) and its dynamics are always in the classical correspondence limit.  Direct participation of multiple nonlinear modes (such as superconducting qubits) in non-reciprocal coupling, as envisioned in chiral quantum optics~\cite{lodahl_chiral_2017}, would lead to novel forms of entanglement stabilization and many-body phases~\cite{stannigel_driven-dissipative_2012, ramos_quantum_2014}.  Our system presents another potential platform to implement this regime in circuit QED in additional to those proposed using dynamic control~\cite{guimond_unidirectional_2020, gheeraert_programmable_2020}.   Strong coupling of Josephson circuits with low-loss non-reciprocal elements can even produce degenerate and protected ground states for robust encoding of qubits~\cite{rymarz_hardware-encoding_2021}.

\begin{acknowledgments}
We thank Juliang Li and Dario Rosenstock for experimental assistance.  This research was supported by U.S. Army Research Office under grants W911-NF-17-1-0469 and W911-NF-19-1-0380.
\end{acknowledgments}

\begin{appendix}
\section{Numerical simulation of the ferrite device}

Finite element analysis software that supports magnetodynamic simulations, such as Ansys HFSS, can be used to simulate our ciculator system with both driven mode and eigenmode solutions. Eigenmode analysis can solve for the frequency and field distributions of our device's eigenmodes, while driven mode analysis reports the S-parameters over frequency.  Here we discuss eigenmode simulations, but driven mode analysis can be carried out similarly. 

It is well known that when the applied field is large and magnetization is saturated along $z$ axis, one can generalize to the whole ferrite the equations of motion derived from the torque experienced by an electron dipole moment under the presence of an applied field. This approach, augmented by the small signal approximation of the Landau-Lifshitz equation of motion, yields the textbook Polder (relative) permeability tensor:
\begin{equation}
[\mu]_z=
\begin{pmatrix}
\mu_r & i\kappa & 0\\
-i\kappa & \mu_r & 0\\
0 & 0 & 1\\
\end{pmatrix}
\end{equation}
where 
$\mu_r = 1 + \frac{\omega_0\omega_m}{\omega_0^2 - \omega^2}$,
$\kappa = \frac{\omega\omega_m}{\omega_0^2 - \omega^2}$,
with $\omega_0 = \gamma\mu_0 H_0$ and $\omega_m = \gamma\mu_0 M_s$ being the internal field strength and saturation magnetization converted to frequencies, respectively. 

The Polder permeability tensor is implimented in HFSS by default to solve for the interaction of a saturated ferrite with an AC microwave field. However, in our experiment we operate the circulator at a low bias field, where the ferrite is not fully saturated.  We adopt a permeability tensor model proposed by Sandy and Green~\cite{green_microwave_1974} for a partially-magnetized ferrite:
\begin{align}
[\mu]_z=
\begin{pmatrix}
\mu_p & i\kappa_p & 0\\
-i\kappa_p & \mu_p & 0\\
0 & 0 & \mu_z\\
\end{pmatrix}
\end{align}
where
\begin{align}
  \mu_p &= \mu_d + (1-\mu_d)\bigg(\frac{|M_p|}{M_s}\bigg)^{3/2}\\
  \kappa_p &= \kappa \bigg(\frac{M_p}{M_s}\bigg)\\
  \mu_z &= \mu_d^{\big(1-\frac{|M_p|}{M_s}\big)^{5/2}}\\
  \mu_d &= \frac{1}{3} + \frac{2}{3}\sqrt{1-\bigg(\frac{\omega_m}{\omega}\bigg)^2}
\end{align}
with $M_p$ being the net magnetization of the partially magnetized ferrite. This model contains functional forms for $\mu_p$ and $\mu_z$ that are purely empirical.  However, the expressions for $\kappa_p$, which dictates the chiral splitting of the circulator modes, and $\mu_d$, which represents the permeability in fully demagnetized state, are well motivated~\cite{schlomann_microwave_1970}.

\begin{figure}[tbp]
    \centering
    \includegraphics[scale=.57]{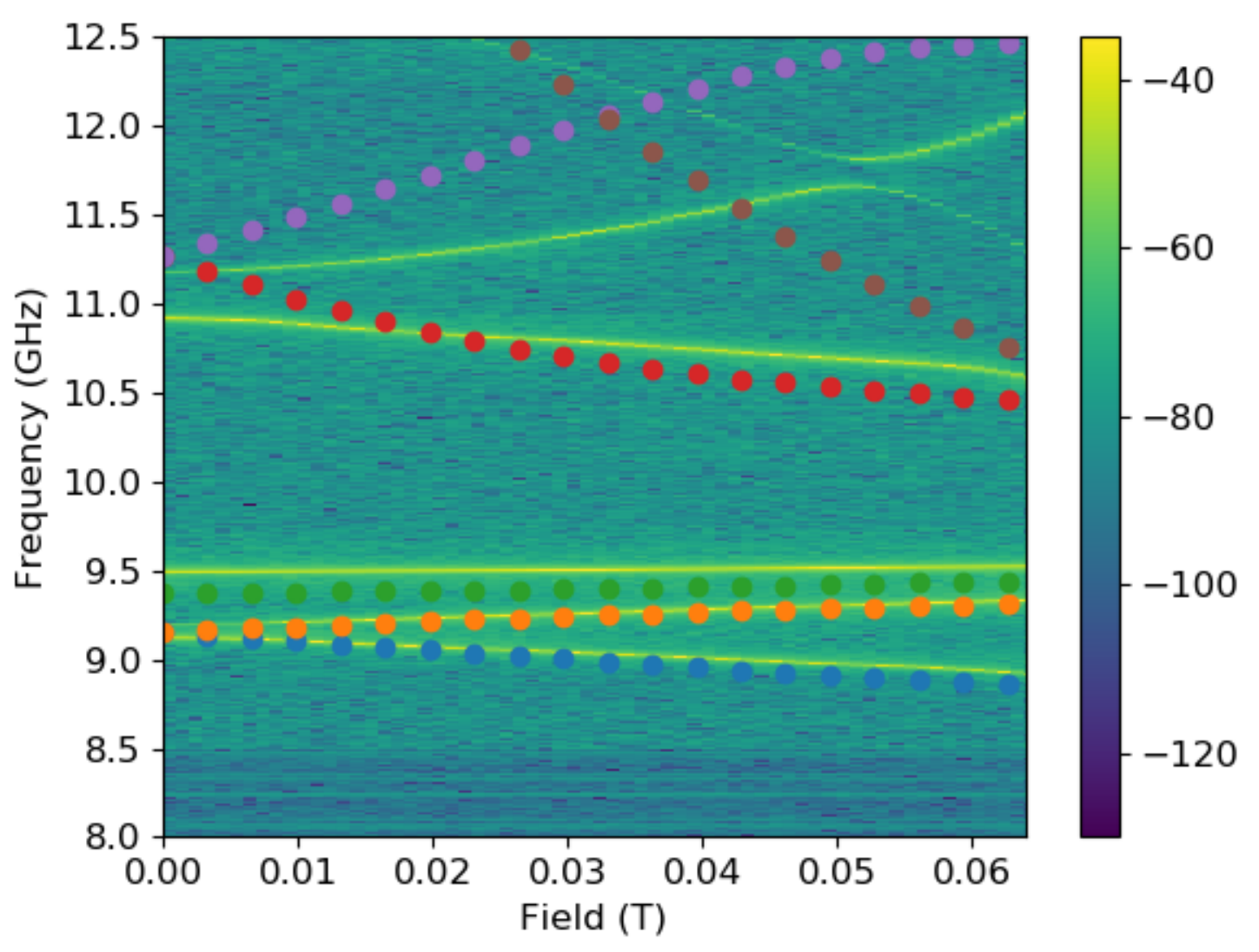}
    \caption{\textbf{Eigenfrequencies of the device from finite-element simulations.}  For the circulator device with WCP, HFSS eigenmode simulation gives the mode frequency over different magnetic fields(dots) and agree semi-quantitatively with experimental data.
    }
    \label{fig:eigenfrequency}
\end{figure}

This model is implemented in simulations by defining materials with the customized permeability tensor as given above.  Whereas HFSS does not by default support eigenmode simulations for a ferrite under a DC bias field, manually defining the permeability tensor components allows us to simulate the circulator's eigenmode structure at any magnetization (bias field)  
as shown in Fig.~\ref{fig:eigenfrequency}.  

The simulation results agree semi-quantitatively with the experimental data in Fig.~\ref{fig:internal_spectrum}(a) with a linear relationship between applied magnetic field and magnetization $M = \mu_0 B/N_z$ mentioned earlier, including a dielectric resonance mode with steep magnetic field dependence that is visible in the experimental data.

Relating to the anisotropy mentioned earlier in section III, the simulation is treating the YIG cylinder as completely isotropic, leading to degenerate Mode x and y at 0 field. To account for the anisotropy, we introduce a general energetic preference along the $x$ axis, thus making the domains of the unsaturated YIG preferentially align along the $x$ axis and breaking the rotational symmetry.

When all domains are oriented 
along the $z$ axis with net magnetization of zero, the permeability is calculated to be 
\begin{equation}
[\mu]_z = \begin{pmatrix}
\mu_\mathrm{eff}&0&0\\
0&\mu_\mathrm{eff}&0\\
0 &0&1\\
\end{pmatrix}
\end{equation}
where $\mu_\mathrm{eff} = \sqrt{\frac{\omega^2 - \omega_m^2 }{\omega^2 }}$. 
To get the permeability matrix for domains align along the $x$ and $y$ axes ($[\mu]_x$, $[\mu]_y$), one can apply a change of coordinates to Eq.~(A7).
The matrix for completely random domain orientations would be an equal average of the three permeability matrices, $[\mu]_x, [\mu]_y, [\mu]_z$~\cite{schlomann_microwave_1970}. Applying a weighted average to the matrices will then allow for representation of an energetic preference, as shown for a preference along the $x$ axis:

\begin{equation}
[\mu] = (\frac{1}{3} +\delta)[\mu]_x + (\frac{1}{3} -\delta)[\mu]_y + \frac{1}{3}[\mu]_z 
\end{equation}

Using $\delta = 0.1$ in Eq.~(A8) gives 260 MHz of splitting between Mode x and Mode y, which is in good agreement with experimental results from Fig.~\ref{fig:internal_spectrum}(b).

\section{Modeling YIG anisotropy in system Hamiltonian}

As mentioned earlier, there is a clear broken rotational symmetry in the $x$-$y$ plane apparent from the splitting in Fig.~2(b). Since the exact origin of the anisotropy is unknown, we will treat it as a general energetic favoring in the x-y plane. As the $\hat{z}$ bias field is increased, the magnetization will align more along $\hat{z}$, making $x$-$y$ plane preferences less impactful. Based off this understanding, we wanted a simple functional form to describe how the effect of this anisotropy decays with an increase in bias field strength that we could use to describe the decay of $\beta$. Since we just want the general form of how the effect of an energetic preference decays over field, the actual form of the energetic preference in the $x$-$y$ plane is not important. We chose to use a toy model of a magnetic domain with a simple Hamiltonian ($H_\mathrm{an}$) with a simple energetic preference given by $K$ along the $x$ axis and a total net magnetic moment $M$:
\begin{equation}
H_\mathrm{an} = -BM \cos(\theta) - K\sin^{2}(\theta)\cos^{2}(\phi)
\end{equation}
To see how the effect of this anisotropy changes as we vary the magnetic field $B$, we utilized classical Boltzmann statistics. We define a partition function:
\begin{equation}
Z =  \int_{0}^{\pi}\int_{0}^{2\pi} e^{-H_\mathrm{an}(\theta,\phi)/(k_{b}T)} \,\sin(\theta)d\theta d\phi \
\end{equation}
so we can calculate the expectation of the magnetic moment direction using Eq.~(B3) for $A = M_{x},M_{y},M_{z}$; $M_{x} = M\sin(\theta)\cos(\phi), M_{y} = M\sin(\theta)\sin(\phi), M_{z} = M\cos(\theta)$.
\begin{equation}
\langle A^{2}\rangle = \int_{0}^{\pi}\int_{0}^{2\pi} A^{2}e^{-H_\mathrm{an}(\theta,\phi)/(k_{b}T)} \,\sin(\theta)d\theta d\phi \ / Z
\end{equation}

We calculate these expectation values numerically, and find that the difference of $\langle M_{x}^{2}\rangle-\langle M_{y}^{2}\rangle$ follows approximately a $\sech(B)$ function. This motivates us to use this simple functional form to model the anisotropy-induced term $\beta$:
\begin{align}
\beta(B) &= \beta_{0} \sech(B/B_{0})
\end{align}
The scaling factor $B_0$ was fit to the $S_{31}$ spectrum giving a value of 18.5 mT. While this is a rather crude phenomenological treatment of the anisotropy, since the detuning of the circulator modes becomes large enough that there is little hybridization with the cavities at relatively small magnetic fields ($\sim$20 mT), the exact dependence on magnetic field becomes less important to understand the non reciprocal dynamics of the cavities.

\section{Two-mode Hamiltonian and gauge symmetry}

We aim to elucidate the non-reciprocity from the Hamiltonian given in Eq.~(4) by reducing it to the form written in Eq.~(9). In order to do this, we adiabatically integrate out the two circulator modes to reduce the Hamiltonian to a simple $2\times2$ matrix ($H'_\mathrm{eff})$ involving only the two cavity modes. The adiabatic elimination is justified due to the large loss rate on the hybridized circulator modes, making their relevant time scales much faster than the time scale set by the coupling parameters to the cavities. The form of the of the effective Hamiltonian is written out in Eq.~(9).
Due to the complicated dependence on the four mode model parameters, we have written simple frequency and loss terms on the diagonal entries and simple non-reciprocal couplings on the off diagonal entries where their explicit values change as a function of magnetic field. The coupling terms ($H_{12},H_{21}$) along with $r = \sqrt{\lvert H_{21}\rvert/\lvert H_{12}\rvert}$ are plotted in Fig.~\ref{fig:2by2 matrix}. The non-reciprocal nature of the system then becomes immediately apparent as the $H_{21}$ and $H_{12}$ Hamiltonian terms are different outside of 0 field, showing a clear directionality in the interaction. 
We can map this Hamiltonian to a reciprocal one using the similarity transformation outlined in Eq.~(C1) with the transformation matrix written in Eq.~(C2). This new reciprocal Hamiltonain is now symmetric under flipping the sign of the magnetic field $H_\mathrm{rec}(B)=H_\mathrm{rec}(-B)$.
\begin{equation}
H'_\mathrm{eff} \rightarrow S H'_\mathrm{eff} S^{-1} \equiv H_\mathrm{rec}
\end{equation}
\begin{equation}
S = \begin{pmatrix}
r^{1/2} & 0\\
0 & r^{-1/2}\\
\end{pmatrix}
\end{equation}
This means that plotting the ratio of $R_{i,n,\mathrm{rec}}$ from $H_\mathrm{rec}$ will always yield 1 for all B values. One can also map the eigenvectors of the original system ($\ket{\psi_{i}}$) to the reciprocal system ($\ket{\psi_{i,\mathrm{rec}}}$) by $\ket{\psi_{Ri,\mathrm{rec}}}$ = $S\ket{\psi_{Ri}}, \ket{\psi_{Li,\mathrm{rec}}}$ = $S^{-1}\ket{\psi_{Li}}$. Starting from the ratio $R_{i,n,\mathrm{rec}}=1$ using the eigenvectors of $H_\mathrm{rec}$, it is then apparent that transforming the eigenvectors back to those of $H'_\mathrm{eff}$ will allow one to simply caluclate $R_{i,n}$.  To illustrate this, we start with the explicit change in components from the transformation of the right and left eigenvectors as written in Eqns.~(C3, C4), we can then substitute these in to the earlier expression for the ratio $R_{i,n}$ and see how the ratio deviates from the reciprocal case of 1, as done in Eq.~(C5) with $i=1$ as an example.
\begin{align}
\ket{\psi_{R,\mathrm{rec}}} = 
\begin{pmatrix}
x\\
y\\
\end{pmatrix}
\xrightarrow[\text{transform}]{\text{similarity}}
\ket{\psi_{R}}=\frac{1}{\sqrt{\frac{|x|^{2}}{r}+|y|^{2}r}}\begin{pmatrix}
\frac{1}{\sqrt{r}}x\\
\sqrt{r}y\\
\end{pmatrix}
\end{align}
\begin{equation}
\ket{\psi_{L,\mathrm{rec}}} = 
\begin{pmatrix}
x^{*}\\
y^{*}\\
\end{pmatrix}
\xrightarrow[\text{transform}]{\text{similarity}}
\ket{\psi_{L}}=\frac{1}{\sqrt{|x|^{2}r+\frac{|y|^{2}}{r}}}\begin{pmatrix}
\sqrt{r}x^{*}\\
\frac{1}{\sqrt{r}}y^{*}\\
\end{pmatrix}
\end{equation}
\begin{equation}
R_{1,n} = \frac{|\bra{n_L}\ket{1}|}{|\bra{1}\ket{n_R}|} = \frac{|r^{1/2}x\sqrt{|x|^{2}r^{-1}+|y|^{2}r|}}{{|r^{-1/2}x\sqrt{|x|^{2}r+|y|^{2}r^{-1}}|}}
\end{equation}
It is important to note two simplifying limits for Eq.~(C5) that the reader may verify themselves, for $x/y \gg r$, $R_{1,n}\approx 1$ and for $y/x \gg r$, $R_{1,n}\approx r^{2}$.

One can use this similarity transformation to understand the qualitative behavior of the disparate amplitude ratios seen in Fig.~\ref{fig:hybrid_spectrum}(f). As mentioned earlier, the amplitude ratio in this case can be roughly approximated as $R_{1,n}$ at $|B| \gtrsim$ 15 mT so we can focus primarily on this ratio to understand the behavior in this field range. At larger fields ($B \gtrsim 20$ mT) Modes a and b are dominated by participation in the bare cavity modes so we can approximate these modes by using the eigenmode values from $H'_\mathrm{eff}$ for the cavity mode components and zeros for the circulator mode components. Under this approximation we can look at the inner products in $R_{1,n}$ just from the components in the eigenmodes of $H'_\mathrm{eff}$.
Mode a is largely dominated by the cavity 1 component with little circulator participation with $|\bra{a_\mathrm{rec}}\ket{1}|/|\bra{a_\mathrm{rec}} \ket{2}| \gg r$ for all values of $r$, thus we can invoke the limit of Eq.(C5) previously mentioned to find the ratio $R_{1,b} \approx 1$ which is what is seen in Fig.~\ref{fig:hybrid_spectrum}(f). The same argument can be made for mode b, but in the opposite limit of $|\bra{b_\mathrm{rec}}\ket{2}|/|\bra{b_\mathrm{rec}} \ket{1}| \gg r$, leading to the other limit of Eq.(C5), making the ratio $R_{1,b} \approx r^{2}$ which can be seen by comparing Fig.~\ref{fig:hybrid_spectrum}(f) with Fig.~\ref{fig:2by2 matrix}.

\end{appendix}

\bibliography{Chen_zotero}

\begin{thebibliography}{54}%
\makeatletter
\providecommand \@ifxundefined [1]{%
 \@ifx{#1\undefined}
}%
\providecommand \@ifnum [1]{%
 \ifnum #1\expandafter \@firstoftwo
 \else \expandafter \@secondoftwo
 \fi
}%
\providecommand \@ifx [1]{%
 \ifx #1\expandafter \@firstoftwo
 \else \expandafter \@secondoftwo
 \fi
}%
\providecommand \natexlab [1]{#1}%
\providecommand \enquote  [1]{``#1''}%
\providecommand \bibnamefont  [1]{#1}%
\providecommand \bibfnamefont [1]{#1}%
\providecommand \citenamefont [1]{#1}%
\providecommand \href@noop [0]{\@secondoftwo}%
\providecommand \href [0]{\begingroup \@sanitize@url \@href}%
\providecommand \@href[1]{\@@startlink{#1}\@@href}%
\providecommand \@@href[1]{\endgroup#1\@@endlink}%
\providecommand \@sanitize@url [0]{\catcode `\\12\catcode `\$12\catcode
  `\&12\catcode `\#12\catcode `\^12\catcode `\_12\catcode `\%12\relax}%
\providecommand \@@startlink[1]{}%
\providecommand \@@endlink[0]{}%
\providecommand \url  [0]{\begingroup\@sanitize@url \@url }%
\providecommand \@url [1]{\endgroup\@href {#1}{\urlprefix }}%
\providecommand \urlprefix  [0]{URL }%
\providecommand \Eprint [0]{\href }%
\providecommand \doibase [0]{http://dx.doi.org/}%
\providecommand \selectlanguage [0]{\@gobble}%
\providecommand \bibinfo  [0]{\@secondoftwo}%
\providecommand \bibfield  [0]{\@secondoftwo}%
\providecommand \translation [1]{[#1]}%
\providecommand \BibitemOpen [0]{}%
\providecommand \bibitemStop [0]{}%
\providecommand \bibitemNoStop [0]{.\EOS\space}%
\providecommand \EOS [0]{\spacefactor3000\relax}%
\providecommand \BibitemShut  [1]{\csname bibitem#1\endcsname}%
\let\auto@bib@innerbib\@empty
\bibitem [{\citenamefont {Kord}\ \emph {et~al.}(2020)\citenamefont {Kord},
  \citenamefont {Sounas},\ and\ \citenamefont {Alù}}]{kord_microwave_2020}%
  \BibitemOpen
  \bibfield  {author} {\bibinfo {author} {\bibfnamefont {A.}~\bibnamefont
  {Kord}}, \bibinfo {author} {\bibfnamefont {D.~L.}\ \bibnamefont {Sounas}}, \
  and\ \bibinfo {author} {\bibfnamefont {A.}~\bibnamefont {Alù}},\ }\href
  {\doibase 10.1109/JPROC.2020.3006041} {\bibfield  {journal} {\bibinfo
  {journal} {Proceedings of the IEEE}\ }\textbf {\bibinfo {volume} {108}},\
  \bibinfo {pages} {1728} (\bibinfo {year} {2020})}\BibitemShut {NoStop}%
\bibitem [{\citenamefont {Devoret}\ and\ \citenamefont
  {Schoelkopf}(2013)}]{devoret_superconducting_2013}%
  \BibitemOpen
  \bibfield  {author} {\bibinfo {author} {\bibfnamefont {M.~H.}\ \bibnamefont
  {Devoret}}\ and\ \bibinfo {author} {\bibfnamefont {R.~J.}\ \bibnamefont
  {Schoelkopf}},\ }\href {\doibase 10.1126/science.1231930} {\bibfield
  {journal} {\bibinfo  {journal} {Science}\ }\textbf {\bibinfo {volume}
  {339}},\ \bibinfo {pages} {1169} (\bibinfo {year} {2013})}\BibitemShut
  {NoStop}%
\bibitem [{\citenamefont {Krantz}\ \emph {et~al.}(2019)\citenamefont {Krantz},
  \citenamefont {Kjaergaard}, \citenamefont {Yan}, \citenamefont {Orlando},
  \citenamefont {Gustavsson},\ and\ \citenamefont
  {Oliver}}]{krantz_quantum_2019}%
  \BibitemOpen
  \bibfield  {author} {\bibinfo {author} {\bibfnamefont {P.}~\bibnamefont
  {Krantz}}, \bibinfo {author} {\bibfnamefont {M.}~\bibnamefont {Kjaergaard}},
  \bibinfo {author} {\bibfnamefont {F.}~\bibnamefont {Yan}}, \bibinfo {author}
  {\bibfnamefont {T.~P.}\ \bibnamefont {Orlando}}, \bibinfo {author}
  {\bibfnamefont {S.}~\bibnamefont {Gustavsson}}, \ and\ \bibinfo {author}
  {\bibfnamefont {W.~D.}\ \bibnamefont {Oliver}},\ }\href {\doibase
  10.1063/1.5089550} {\bibfield  {journal} {\bibinfo  {journal} {Applied
  Physics Reviews}\ }\textbf {\bibinfo {volume} {6}},\ \bibinfo {pages}
  {021318} (\bibinfo {year} {2019})}\BibitemShut {NoStop}%
\bibitem [{\citenamefont {Kurpiers}\ \emph {et~al.}(2018)\citenamefont
  {Kurpiers}, \citenamefont {Magnard}, \citenamefont {Walter}, \citenamefont
  {Royer}, \citenamefont {Pechal}, \citenamefont {Heinsoo}, \citenamefont
  {Salathé}, \citenamefont {Akin}, \citenamefont {Storz}, \citenamefont
  {Besse}, \citenamefont {Gasparinetti}, \citenamefont {Blais},\ and\
  \citenamefont {Wallraff}}]{kurpiers_deterministic_2018}%
  \BibitemOpen
  \bibfield  {author} {\bibinfo {author} {\bibfnamefont {P.}~\bibnamefont
  {Kurpiers}}, \bibinfo {author} {\bibfnamefont {P.}~\bibnamefont {Magnard}},
  \bibinfo {author} {\bibfnamefont {T.}~\bibnamefont {Walter}}, \bibinfo
  {author} {\bibfnamefont {B.}~\bibnamefont {Royer}}, \bibinfo {author}
  {\bibfnamefont {M.}~\bibnamefont {Pechal}}, \bibinfo {author} {\bibfnamefont
  {J.}~\bibnamefont {Heinsoo}}, \bibinfo {author} {\bibfnamefont
  {Y.}~\bibnamefont {Salathé}}, \bibinfo {author} {\bibfnamefont
  {A.}~\bibnamefont {Akin}}, \bibinfo {author} {\bibfnamefont {S.}~\bibnamefont
  {Storz}}, \bibinfo {author} {\bibfnamefont {J.-C.}\ \bibnamefont {Besse}},
  \bibinfo {author} {\bibfnamefont {S.}~\bibnamefont {Gasparinetti}}, \bibinfo
  {author} {\bibfnamefont {A.}~\bibnamefont {Blais}}, \ and\ \bibinfo {author}
  {\bibfnamefont {A.}~\bibnamefont {Wallraff}},\ }\href {\doibase
  10.1038/s41586-018-0195-y} {\bibfield  {journal} {\bibinfo  {journal}
  {Nature}\ }\textbf {\bibinfo {volume} {558}},\ \bibinfo {pages} {264}
  (\bibinfo {year} {2018})}\BibitemShut {NoStop}%
\bibitem [{\citenamefont {Axline}\ \emph {et~al.}(2018)\citenamefont {Axline},
  \citenamefont {Burkhart}, \citenamefont {Pfaff}, \citenamefont {Zhang},
  \citenamefont {Chou}, \citenamefont {Campagne-Ibarcq}, \citenamefont
  {Reinhold}, \citenamefont {Frunzio}, \citenamefont {Girvin}, \citenamefont
  {Jiang}, \citenamefont {Devoret},\ and\ \citenamefont
  {Schoelkopf}}]{axline_-demand_2018}%
  \BibitemOpen
  \bibfield  {author} {\bibinfo {author} {\bibfnamefont {C.~J.}\ \bibnamefont
  {Axline}}, \bibinfo {author} {\bibfnamefont {L.~D.}\ \bibnamefont
  {Burkhart}}, \bibinfo {author} {\bibfnamefont {W.}~\bibnamefont {Pfaff}},
  \bibinfo {author} {\bibfnamefont {M.}~\bibnamefont {Zhang}}, \bibinfo
  {author} {\bibfnamefont {K.}~\bibnamefont {Chou}}, \bibinfo {author}
  {\bibfnamefont {P.}~\bibnamefont {Campagne-Ibarcq}}, \bibinfo {author}
  {\bibfnamefont {P.}~\bibnamefont {Reinhold}}, \bibinfo {author}
  {\bibfnamefont {L.}~\bibnamefont {Frunzio}}, \bibinfo {author} {\bibfnamefont
  {S.~M.}\ \bibnamefont {Girvin}}, \bibinfo {author} {\bibfnamefont
  {L.}~\bibnamefont {Jiang}}, \bibinfo {author} {\bibfnamefont {M.~H.}\
  \bibnamefont {Devoret}}, \ and\ \bibinfo {author} {\bibfnamefont {R.~J.}\
  \bibnamefont {Schoelkopf}},\ }\href {\doibase 10.1038/s41567-018-0115-y}
  {\bibfield  {journal} {\bibinfo  {journal} {Nature Physics}\ }\textbf
  {\bibinfo {volume} {14}},\ \bibinfo {pages} {705} (\bibinfo {year}
  {2018})}\BibitemShut {NoStop}%
\bibitem [{\citenamefont {Vool}\ and\ \citenamefont
  {Devoret}(2017)}]{vool_introduction_2017}%
  \BibitemOpen
  \bibfield  {author} {\bibinfo {author} {\bibfnamefont {U.}~\bibnamefont
  {Vool}}\ and\ \bibinfo {author} {\bibfnamefont {M.}~\bibnamefont {Devoret}},\
  }\href {\doibase https://doi.org/10.1002/cta.2359} {\bibfield  {journal}
  {\bibinfo  {journal} {International Journal of Circuit Theory and
  Applications}\ }\textbf {\bibinfo {volume} {45}},\ \bibinfo {pages} {897}
  (\bibinfo {year} {2017})}\BibitemShut {NoStop}%
\bibitem [{\citenamefont {Lachance-Quirion}\ \emph {et~al.}(2019)\citenamefont
  {Lachance-Quirion}, \citenamefont {Tabuchi}, \citenamefont {Gloppe},
  \citenamefont {Usami},\ and\ \citenamefont
  {Nakamura}}]{lachance-quirion_hybrid_2019}%
  \BibitemOpen
  \bibfield  {author} {\bibinfo {author} {\bibfnamefont {D.}~\bibnamefont
  {Lachance-Quirion}}, \bibinfo {author} {\bibfnamefont {Y.}~\bibnamefont
  {Tabuchi}}, \bibinfo {author} {\bibfnamefont {A.}~\bibnamefont {Gloppe}},
  \bibinfo {author} {\bibfnamefont {K.}~\bibnamefont {Usami}}, \ and\ \bibinfo
  {author} {\bibfnamefont {Y.}~\bibnamefont {Nakamura}},\ }\href {\doibase
  10.7567/1882-0786/ab248d} {\bibfield  {journal} {\bibinfo  {journal} {Applied
  Physics Express}\ }\textbf {\bibinfo {volume} {12}},\ \bibinfo {pages}
  {070101} (\bibinfo {year} {2019})}\BibitemShut {NoStop}%
\bibitem [{\citenamefont {Awschalom}\ \emph {et~al.}(2021)\citenamefont
  {Awschalom}, \citenamefont {Du}, \citenamefont {He}, \citenamefont
  {Heremans}, \citenamefont {Hoffmann}, \citenamefont {Hou}, \citenamefont
  {Kurebayashi}, \citenamefont {Li}, \citenamefont {Liu}, \citenamefont
  {Novosad}, \citenamefont {Sklenar}, \citenamefont {Sullivan}, \citenamefont
  {Sun}, \citenamefont {Tang}, \citenamefont {Tiberkevich}, \citenamefont
  {Trevillian}, \citenamefont {Tsen}, \citenamefont {Weiss}, \citenamefont
  {Zhang}, \citenamefont {Zhang}, \citenamefont {Zhao},\ and\ \citenamefont
  {Zollitsch}}]{awschalom_quantum_2021}%
  \BibitemOpen
  \bibfield  {author} {\bibinfo {author} {\bibfnamefont {D.~D.}\ \bibnamefont
  {Awschalom}}, \bibinfo {author} {\bibfnamefont {C.~H.~R.}\ \bibnamefont
  {Du}}, \bibinfo {author} {\bibfnamefont {R.}~\bibnamefont {He}}, \bibinfo
  {author} {\bibfnamefont {F.~J.}\ \bibnamefont {Heremans}}, \bibinfo {author}
  {\bibfnamefont {A.}~\bibnamefont {Hoffmann}}, \bibinfo {author}
  {\bibfnamefont {J.~T.}\ \bibnamefont {Hou}}, \bibinfo {author} {\bibfnamefont
  {H.}~\bibnamefont {Kurebayashi}}, \bibinfo {author} {\bibfnamefont
  {Y.}~\bibnamefont {Li}}, \bibinfo {author} {\bibfnamefont {L.}~\bibnamefont
  {Liu}}, \bibinfo {author} {\bibfnamefont {V.}~\bibnamefont {Novosad}},
  \bibinfo {author} {\bibfnamefont {J.}~\bibnamefont {Sklenar}}, \bibinfo
  {author} {\bibfnamefont {S.~E.}\ \bibnamefont {Sullivan}}, \bibinfo {author}
  {\bibfnamefont {D.}~\bibnamefont {Sun}}, \bibinfo {author} {\bibfnamefont
  {H.}~\bibnamefont {Tang}}, \bibinfo {author} {\bibfnamefont {V.}~\bibnamefont
  {Tiberkevich}}, \bibinfo {author} {\bibfnamefont {C.}~\bibnamefont
  {Trevillian}}, \bibinfo {author} {\bibfnamefont {A.~W.}\ \bibnamefont
  {Tsen}}, \bibinfo {author} {\bibfnamefont {L.~R.}\ \bibnamefont {Weiss}},
  \bibinfo {author} {\bibfnamefont {W.}~\bibnamefont {Zhang}}, \bibinfo
  {author} {\bibfnamefont {X.}~\bibnamefont {Zhang}}, \bibinfo {author}
  {\bibfnamefont {L.}~\bibnamefont {Zhao}}, \ and\ \bibinfo {author}
  {\bibfnamefont {C.~W.}\ \bibnamefont {Zollitsch}},\ }\href
  {http://arxiv.org/abs/2102.03222} {\bibfield  {journal} {\bibinfo  {journal}
  {arXiv:2102.03222}\ } (\bibinfo {year} {2021})}\BibitemShut {NoStop}%
\bibitem [{\citenamefont {Huebl}\ \emph {et~al.}(2013)\citenamefont {Huebl},
  \citenamefont {Zollitsch}, \citenamefont {Lotze}, \citenamefont {Hocke},
  \citenamefont {Greifenstein}, \citenamefont {Marx}, \citenamefont {Gross},\
  and\ \citenamefont {Goennenwein}}]{huebl_high_2013}%
  \BibitemOpen
  \bibfield  {author} {\bibinfo {author} {\bibfnamefont {H.}~\bibnamefont
  {Huebl}}, \bibinfo {author} {\bibfnamefont {C.~W.}\ \bibnamefont
  {Zollitsch}}, \bibinfo {author} {\bibfnamefont {J.}~\bibnamefont {Lotze}},
  \bibinfo {author} {\bibfnamefont {F.}~\bibnamefont {Hocke}}, \bibinfo
  {author} {\bibfnamefont {M.}~\bibnamefont {Greifenstein}}, \bibinfo {author}
  {\bibfnamefont {A.}~\bibnamefont {Marx}}, \bibinfo {author} {\bibfnamefont
  {R.}~\bibnamefont {Gross}}, \ and\ \bibinfo {author} {\bibfnamefont
  {S.~T.~B.}\ \bibnamefont {Goennenwein}},\ }\href {\doibase
  10.1103/PhysRevLett.111.127003} {\bibfield  {journal} {\bibinfo  {journal}
  {Physical Review Letters}\ }\textbf {\bibinfo {volume} {111}},\ \bibinfo
  {pages} {127003} (\bibinfo {year} {2013})}\BibitemShut {NoStop}%
\bibitem [{\citenamefont {Tabuchi}\ \emph {et~al.}(2014)\citenamefont
  {Tabuchi}, \citenamefont {Ishino}, \citenamefont {Ishikawa}, \citenamefont
  {Yamazaki}, \citenamefont {Usami},\ and\ \citenamefont
  {Nakamura}}]{tabuchi_hybridizing_2014}%
  \BibitemOpen
  \bibfield  {author} {\bibinfo {author} {\bibfnamefont {Y.}~\bibnamefont
  {Tabuchi}}, \bibinfo {author} {\bibfnamefont {S.}~\bibnamefont {Ishino}},
  \bibinfo {author} {\bibfnamefont {T.}~\bibnamefont {Ishikawa}}, \bibinfo
  {author} {\bibfnamefont {R.}~\bibnamefont {Yamazaki}}, \bibinfo {author}
  {\bibfnamefont {K.}~\bibnamefont {Usami}}, \ and\ \bibinfo {author}
  {\bibfnamefont {Y.}~\bibnamefont {Nakamura}},\ }\href {\doibase
  10.1103/PhysRevLett.113.083603} {\bibfield  {journal} {\bibinfo  {journal}
  {Physical Review Letters}\ }\textbf {\bibinfo {volume} {113}},\ \bibinfo
  {pages} {083603} (\bibinfo {year} {2014})}\BibitemShut {NoStop}%
\bibitem [{\citenamefont {Zhang}\ \emph {et~al.}(2014)\citenamefont {Zhang},
  \citenamefont {Zou}, \citenamefont {Jiang},\ and\ \citenamefont
  {Tang}}]{zhang_strongly_2014}%
  \BibitemOpen
  \bibfield  {author} {\bibinfo {author} {\bibfnamefont {X.}~\bibnamefont
  {Zhang}}, \bibinfo {author} {\bibfnamefont {C.-L.}\ \bibnamefont {Zou}},
  \bibinfo {author} {\bibfnamefont {L.}~\bibnamefont {Jiang}}, \ and\ \bibinfo
  {author} {\bibfnamefont {H.~X.}\ \bibnamefont {Tang}},\ }\href {\doibase
  10.1103/PhysRevLett.113.156401} {\bibfield  {journal} {\bibinfo  {journal}
  {Physical Review Letters}\ }\textbf {\bibinfo {volume} {113}},\ \bibinfo
  {pages} {156401} (\bibinfo {year} {2014})}\BibitemShut {NoStop}%
\bibitem [{\citenamefont {Tabuchi}\ \emph {et~al.}(2015)\citenamefont
  {Tabuchi}, \citenamefont {Ishino}, \citenamefont {Noguchi}, \citenamefont
  {Ishikawa}, \citenamefont {Yamazaki}, \citenamefont {Usami},\ and\
  \citenamefont {Nakamura}}]{tabuchi_coherent_2015}%
  \BibitemOpen
  \bibfield  {author} {\bibinfo {author} {\bibfnamefont {Y.}~\bibnamefont
  {Tabuchi}}, \bibinfo {author} {\bibfnamefont {S.}~\bibnamefont {Ishino}},
  \bibinfo {author} {\bibfnamefont {A.}~\bibnamefont {Noguchi}}, \bibinfo
  {author} {\bibfnamefont {T.}~\bibnamefont {Ishikawa}}, \bibinfo {author}
  {\bibfnamefont {R.}~\bibnamefont {Yamazaki}}, \bibinfo {author}
  {\bibfnamefont {K.}~\bibnamefont {Usami}}, \ and\ \bibinfo {author}
  {\bibfnamefont {Y.}~\bibnamefont {Nakamura}},\ }\href {\doibase
  10.1126/science.aaa3693} {\bibfield  {journal} {\bibinfo  {journal}
  {Science}\ }\textbf {\bibinfo {volume} {349}},\ \bibinfo {pages} {405}
  (\bibinfo {year} {2015})}\BibitemShut {NoStop}%
\bibitem [{\citenamefont {Lachance-Quirion}\ \emph {et~al.}(2020)\citenamefont
  {Lachance-Quirion}, \citenamefont {Wolski}, \citenamefont {Tabuchi},
  \citenamefont {Kono}, \citenamefont {Usami},\ and\ \citenamefont
  {Nakamura}}]{lachance-quirion_entanglement-based_2020}%
  \BibitemOpen
  \bibfield  {author} {\bibinfo {author} {\bibfnamefont {D.}~\bibnamefont
  {Lachance-Quirion}}, \bibinfo {author} {\bibfnamefont {S.~P.}\ \bibnamefont
  {Wolski}}, \bibinfo {author} {\bibfnamefont {Y.}~\bibnamefont {Tabuchi}},
  \bibinfo {author} {\bibfnamefont {S.}~\bibnamefont {Kono}}, \bibinfo {author}
  {\bibfnamefont {K.}~\bibnamefont {Usami}}, \ and\ \bibinfo {author}
  {\bibfnamefont {Y.}~\bibnamefont {Nakamura}},\ }\href {\doibase
  10.1126/science.aaz9236} {\bibfield  {journal} {\bibinfo  {journal}
  {Science}\ }\textbf {\bibinfo {volume} {367}},\ \bibinfo {pages} {425}
  (\bibinfo {year} {2020})}\BibitemShut {NoStop}%
\bibitem [{\citenamefont {Hou}\ and\ \citenamefont
  {Liu}(2019)}]{hou_strong_2019}%
  \BibitemOpen
  \bibfield  {author} {\bibinfo {author} {\bibfnamefont {J.~T.}\ \bibnamefont
  {Hou}}\ and\ \bibinfo {author} {\bibfnamefont {L.}~\bibnamefont {Liu}},\
  }\href {\doibase 10.1103/PhysRevLett.123.107702} {\bibfield  {journal}
  {\bibinfo  {journal} {Physical Review Letters}\ }\textbf {\bibinfo {volume}
  {123}},\ \bibinfo {pages} {107702} (\bibinfo {year} {2019})}\BibitemShut
  {NoStop}%
\bibitem [{\citenamefont {Golovchanskiy}\ \emph {et~al.}(2021)\citenamefont
  {Golovchanskiy}, \citenamefont {Abramov}, \citenamefont {Stolyarov},
  \citenamefont {Weides}, \citenamefont {Ryazanov}, \citenamefont {Golubov},
  \citenamefont {Ustinov},\ and\ \citenamefont
  {Kupriyanov}}]{golovchanskiy_ultrastrong_2021}%
  \BibitemOpen
  \bibfield  {author} {\bibinfo {author} {\bibfnamefont {I.~A.}\ \bibnamefont
  {Golovchanskiy}}, \bibinfo {author} {\bibfnamefont {N.~N.}\ \bibnamefont
  {Abramov}}, \bibinfo {author} {\bibfnamefont {V.~S.}\ \bibnamefont
  {Stolyarov}}, \bibinfo {author} {\bibfnamefont {M.}~\bibnamefont {Weides}},
  \bibinfo {author} {\bibfnamefont {V.~V.}\ \bibnamefont {Ryazanov}}, \bibinfo
  {author} {\bibfnamefont {A.~A.}\ \bibnamefont {Golubov}}, \bibinfo {author}
  {\bibfnamefont {A.~V.}\ \bibnamefont {Ustinov}}, \ and\ \bibinfo {author}
  {\bibfnamefont {M.~Y.}\ \bibnamefont {Kupriyanov}},\ }\href {\doibase
  10.1126/sciadv.abe8638} {\bibfield  {journal} {\bibinfo  {journal} {Science
  Advances}\ }\textbf {\bibinfo {volume} {7}},\ \bibinfo {pages} {eabe8638}
  (\bibinfo {year} {2021})}\BibitemShut {NoStop}%
\bibitem [{\citenamefont {Heyroth}\ \emph {et~al.}(2019)\citenamefont
  {Heyroth}, \citenamefont {Hauser}, \citenamefont {Trempler}, \citenamefont
  {Geyer}, \citenamefont {Syrowatka}, \citenamefont {Dreyer}, \citenamefont
  {Ebbinghaus}, \citenamefont {Woltersdorf},\ and\ \citenamefont
  {Schmidt}}]{heyroth_monocrystalline_2019}%
  \BibitemOpen
  \bibfield  {author} {\bibinfo {author} {\bibfnamefont {F.}~\bibnamefont
  {Heyroth}}, \bibinfo {author} {\bibfnamefont {C.}~\bibnamefont {Hauser}},
  \bibinfo {author} {\bibfnamefont {P.}~\bibnamefont {Trempler}}, \bibinfo
  {author} {\bibfnamefont {P.}~\bibnamefont {Geyer}}, \bibinfo {author}
  {\bibfnamefont {F.}~\bibnamefont {Syrowatka}}, \bibinfo {author}
  {\bibfnamefont {R.}~\bibnamefont {Dreyer}}, \bibinfo {author} {\bibfnamefont
  {S.~G.}\ \bibnamefont {Ebbinghaus}}, \bibinfo {author} {\bibfnamefont
  {G.}~\bibnamefont {Woltersdorf}}, \ and\ \bibinfo {author} {\bibfnamefont
  {G.}~\bibnamefont {Schmidt}},\ }\href {\doibase
  10.1103/PhysRevApplied.12.054031} {\bibfield  {journal} {\bibinfo  {journal}
  {Physical Review Applied}\ }\textbf {\bibinfo {volume} {12}},\ \bibinfo
  {pages} {054031} (\bibinfo {year} {2019})}\BibitemShut {NoStop}%
\bibitem [{\citenamefont {Boventer}\ \emph {et~al.}(2018)\citenamefont
  {Boventer}, \citenamefont {Pfirrmann}, \citenamefont {Krause}, \citenamefont
  {Schön}, \citenamefont {Klaui},\ and\ \citenamefont
  {Weides}}]{boventer_complex_2018}%
  \BibitemOpen
  \bibfield  {author} {\bibinfo {author} {\bibfnamefont {I.}~\bibnamefont
  {Boventer}}, \bibinfo {author} {\bibfnamefont {M.}~\bibnamefont {Pfirrmann}},
  \bibinfo {author} {\bibfnamefont {J.}~\bibnamefont {Krause}}, \bibinfo
  {author} {\bibfnamefont {Y.}~\bibnamefont {Schön}}, \bibinfo {author}
  {\bibfnamefont {M.}~\bibnamefont {Klaui}}, \ and\ \bibinfo {author}
  {\bibfnamefont {M.}~\bibnamefont {Weides}},\ }\href {\doibase
  10.1103/PhysRevB.97.184420} {\bibfield  {journal} {\bibinfo  {journal}
  {Physical Review B}\ }\textbf {\bibinfo {volume} {97}},\ \bibinfo {pages}
  {184420} (\bibinfo {year} {2018})}\BibitemShut {NoStop}%
\bibitem [{\citenamefont {Zhang}\ \emph {et~al.}(2017)\citenamefont {Zhang},
  \citenamefont {Luo}, \citenamefont {Wang}, \citenamefont {Li},\ and\
  \citenamefont {You}}]{zhang_observation_2017}%
  \BibitemOpen
  \bibfield  {author} {\bibinfo {author} {\bibfnamefont {D.}~\bibnamefont
  {Zhang}}, \bibinfo {author} {\bibfnamefont {X.-Q.}\ \bibnamefont {Luo}},
  \bibinfo {author} {\bibfnamefont {Y.-P.}\ \bibnamefont {Wang}}, \bibinfo
  {author} {\bibfnamefont {T.-F.}\ \bibnamefont {Li}}, \ and\ \bibinfo {author}
  {\bibfnamefont {J.~Q.}\ \bibnamefont {You}},\ }\href {\doibase
  10.1038/s41467-017-01634-w} {\bibfield  {journal} {\bibinfo  {journal}
  {Nature Communications}\ }\textbf {\bibinfo {volume} {8}},\ \bibinfo {pages}
  {1368} (\bibinfo {year} {2017})}\BibitemShut {NoStop}%
\bibitem [{\citenamefont {Anderson}\ \emph {et~al.}(2016)\citenamefont
  {Anderson}, \citenamefont {Ma}, \citenamefont {Owens}, \citenamefont
  {Schuster},\ and\ \citenamefont {Simon}}]{anderson_engineering_2016}%
  \BibitemOpen
  \bibfield  {author} {\bibinfo {author} {\bibfnamefont {B.~M.}\ \bibnamefont
  {Anderson}}, \bibinfo {author} {\bibfnamefont {R.}~\bibnamefont {Ma}},
  \bibinfo {author} {\bibfnamefont {C.}~\bibnamefont {Owens}}, \bibinfo
  {author} {\bibfnamefont {D.~I.}\ \bibnamefont {Schuster}}, \ and\ \bibinfo
  {author} {\bibfnamefont {J.}~\bibnamefont {Simon}},\ }\href {\doibase
  10.1103/PhysRevX.6.041043} {\bibfield  {journal} {\bibinfo  {journal}
  {Physical Review X}\ }\textbf {\bibinfo {volume} {6}},\ \bibinfo {pages}
  {041043} (\bibinfo {year} {2016})}\BibitemShut {NoStop}%
\bibitem [{\citenamefont {Owens}\ \emph {et~al.}(2018)\citenamefont {Owens},
  \citenamefont {LaChapelle}, \citenamefont {Saxberg}, \citenamefont
  {Anderson}, \citenamefont {Ma}, \citenamefont {Simon},\ and\ \citenamefont
  {Schuster}}]{owens_quarter-flux_2018}%
  \BibitemOpen
  \bibfield  {author} {\bibinfo {author} {\bibfnamefont {C.}~\bibnamefont
  {Owens}}, \bibinfo {author} {\bibfnamefont {A.}~\bibnamefont {LaChapelle}},
  \bibinfo {author} {\bibfnamefont {B.}~\bibnamefont {Saxberg}}, \bibinfo
  {author} {\bibfnamefont {B.~M.}\ \bibnamefont {Anderson}}, \bibinfo {author}
  {\bibfnamefont {R.}~\bibnamefont {Ma}}, \bibinfo {author} {\bibfnamefont
  {J.}~\bibnamefont {Simon}}, \ and\ \bibinfo {author} {\bibfnamefont {D.~I.}\
  \bibnamefont {Schuster}},\ }\href {\doibase 10.1103/PhysRevA.97.013818}
  {\bibfield  {journal} {\bibinfo  {journal} {Physical Review A}\ }\textbf
  {\bibinfo {volume} {97}},\ \bibinfo {pages} {013818} (\bibinfo {year}
  {2018})}\BibitemShut {NoStop}%
\bibitem [{\citenamefont {Zhang}\ \emph {et~al.}(2020)\citenamefont {Zhang},
  \citenamefont {Galda}, \citenamefont {Han}, \citenamefont {Jin},\ and\
  \citenamefont {Vinokur}}]{zhang_broadband_2020}%
  \BibitemOpen
  \bibfield  {author} {\bibinfo {author} {\bibfnamefont {X.}~\bibnamefont
  {Zhang}}, \bibinfo {author} {\bibfnamefont {A.}~\bibnamefont {Galda}},
  \bibinfo {author} {\bibfnamefont {X.}~\bibnamefont {Han}}, \bibinfo {author}
  {\bibfnamefont {D.}~\bibnamefont {Jin}}, \ and\ \bibinfo {author}
  {\bibfnamefont {V.~M.}\ \bibnamefont {Vinokur}},\ }\href {\doibase
  10.1103/PhysRevApplied.13.044039} {\bibfield  {journal} {\bibinfo  {journal}
  {Physical Review Applied}\ }\textbf {\bibinfo {volume} {13}},\ \bibinfo
  {pages} {044039} (\bibinfo {year} {2020})}\BibitemShut {NoStop}%
\bibitem [{\citenamefont {Fay}\ and\ \citenamefont
  {Comstock}(1965)}]{fay_operation_1965}%
  \BibitemOpen
  \bibfield  {author} {\bibinfo {author} {\bibfnamefont {C.~E.}\ \bibnamefont
  {Fay}}\ and\ \bibinfo {author} {\bibfnamefont {R.~L.}\ \bibnamefont
  {Comstock}},\ }\href {\doibase 10.1109/TMTT.1965.1125923} {\bibfield
  {journal} {\bibinfo  {journal} {IEEE Transactions on Microwave Theory and
  Techniques}\ }\textbf {\bibinfo {volume} {13}},\ \bibinfo {pages} {15}
  (\bibinfo {year} {1965})}\BibitemShut {NoStop}%
\bibitem [{\citenamefont {Heiss}(2012)}]{heiss_physics_2012}%
  \BibitemOpen
  \bibfield  {author} {\bibinfo {author} {\bibfnamefont {W.~D.}\ \bibnamefont
  {Heiss}},\ }\href {\doibase 10.1088/1751-8113/45/44/444016} {\ \textbf
  {\bibinfo {volume} {45}},\ \bibinfo {pages} {444016} (\bibinfo {year}
  {2012})}\BibitemShut {NoStop}%
\bibitem [{\citenamefont {Özdemir}\ \emph {et~al.}(2019)\citenamefont
  {Özdemir}, \citenamefont {Rotter}, \citenamefont {Nori},\ and\ \citenamefont
  {Yang}}]{ozdemir_paritytime_2019}%
  \BibitemOpen
  \bibfield  {author} {\bibinfo {author} {\bibfnamefont {S.~K.}\ \bibnamefont
  {Özdemir}}, \bibinfo {author} {\bibfnamefont {S.}~\bibnamefont {Rotter}},
  \bibinfo {author} {\bibfnamefont {F.}~\bibnamefont {Nori}}, \ and\ \bibinfo
  {author} {\bibfnamefont {L.}~\bibnamefont {Yang}},\ }\href {\doibase
  10.1038/s41563-019-0304-9} {\bibfield  {journal} {\bibinfo  {journal} {Nature
  Materials}\ }\textbf {\bibinfo {volume} {18}},\ \bibinfo {pages} {783}
  (\bibinfo {year} {2019})}\BibitemShut {NoStop}%
\bibitem [{\citenamefont {Hatano}\ and\ \citenamefont
  {Nelson}(1997)}]{hatano_vortex_1997}%
  \BibitemOpen
  \bibfield  {author} {\bibinfo {author} {\bibfnamefont {N.}~\bibnamefont
  {Hatano}}\ and\ \bibinfo {author} {\bibfnamefont {D.~R.}\ \bibnamefont
  {Nelson}},\ }\href {\doibase 10.1103/PhysRevB.56.8651} {\bibfield  {journal}
  {\bibinfo  {journal} {Physical Review B}\ }\textbf {\bibinfo {volume} {56}},\
  \bibinfo {pages} {8651} (\bibinfo {year} {1997})}\BibitemShut {NoStop}%
\bibitem [{\citenamefont {Yao}\ and\ \citenamefont
  {Wang}(2018)}]{yao_edge_2018}%
  \BibitemOpen
  \bibfield  {author} {\bibinfo {author} {\bibfnamefont {S.}~\bibnamefont
  {Yao}}\ and\ \bibinfo {author} {\bibfnamefont {Z.}~\bibnamefont {Wang}},\
  }\href {\doibase 10.1103/PhysRevLett.121.086803} {\bibfield  {journal}
  {\bibinfo  {journal} {Physical Review Letters}\ }\textbf {\bibinfo {volume}
  {121}},\ \bibinfo {pages} {086803} (\bibinfo {year} {2018})}\BibitemShut
  {NoStop}%
\bibitem [{\citenamefont {McDonald}\ \emph {et~al.}(2018)\citenamefont
  {McDonald}, \citenamefont {Pereg-Barnea},\ and\ \citenamefont
  {Clerk}}]{mcdonald_phase-dependent_2018}%
  \BibitemOpen
  \bibfield  {author} {\bibinfo {author} {\bibfnamefont {A.}~\bibnamefont
  {McDonald}}, \bibinfo {author} {\bibfnamefont {T.}~\bibnamefont
  {Pereg-Barnea}}, \ and\ \bibinfo {author} {\bibfnamefont {A.~A.}\
  \bibnamefont {Clerk}},\ }\href {\doibase 10.1103/PhysRevX.8.041031}
  {\bibfield  {journal} {\bibinfo  {journal} {Physical Review X}\ }\textbf
  {\bibinfo {volume} {8}},\ \bibinfo {pages} {041031} (\bibinfo {year}
  {2018})}\BibitemShut {NoStop}%
\bibitem [{\citenamefont {Stannigel}\ \emph {et~al.}(2012)\citenamefont
  {Stannigel}, \citenamefont {Rabl},\ and\ \citenamefont
  {Zoller}}]{stannigel_driven-dissipative_2012}%
  \BibitemOpen
  \bibfield  {author} {\bibinfo {author} {\bibfnamefont {K.}~\bibnamefont
  {Stannigel}}, \bibinfo {author} {\bibfnamefont {P.}~\bibnamefont {Rabl}}, \
  and\ \bibinfo {author} {\bibfnamefont {P.}~\bibnamefont {Zoller}},\ }\href
  {\doibase 10.1088/1367-2630/14/6/063014} {\bibfield  {journal} {\bibinfo
  {journal} {New Journal of Physics}\ }\textbf {\bibinfo {volume} {14}},\
  \bibinfo {pages} {063014} (\bibinfo {year} {2012})}\BibitemShut {NoStop}%
\bibitem [{\citenamefont {Lodahl}\ \emph {et~al.}(2017)\citenamefont {Lodahl},
  \citenamefont {Mahmoodian}, \citenamefont {Stobbe}, \citenamefont
  {Rauschenbeutel}, \citenamefont {Schneeweiss}, \citenamefont {Volz},
  \citenamefont {Pichler},\ and\ \citenamefont {Zoller}}]{lodahl_chiral_2017}%
  \BibitemOpen
  \bibfield  {author} {\bibinfo {author} {\bibfnamefont {P.}~\bibnamefont
  {Lodahl}}, \bibinfo {author} {\bibfnamefont {S.}~\bibnamefont {Mahmoodian}},
  \bibinfo {author} {\bibfnamefont {S.}~\bibnamefont {Stobbe}}, \bibinfo
  {author} {\bibfnamefont {A.}~\bibnamefont {Rauschenbeutel}}, \bibinfo
  {author} {\bibfnamefont {P.}~\bibnamefont {Schneeweiss}}, \bibinfo {author}
  {\bibfnamefont {J.}~\bibnamefont {Volz}}, \bibinfo {author} {\bibfnamefont
  {H.}~\bibnamefont {Pichler}}, \ and\ \bibinfo {author} {\bibfnamefont
  {P.}~\bibnamefont {Zoller}},\ }\href {\doibase 10.1038/nature21037}
  {\bibfield  {journal} {\bibinfo  {journal} {Nature}\ }\textbf {\bibinfo
  {volume} {541}},\ \bibinfo {pages} {473} (\bibinfo {year}
  {2017})}\BibitemShut {NoStop}%
\bibitem [{\citenamefont {Chapman}\ \emph {et~al.}(2017)\citenamefont
  {Chapman}, \citenamefont {Rosenthal}, \citenamefont {Kerckhoff},
  \citenamefont {Moores}, \citenamefont {Vale}, \citenamefont {Mates},
  \citenamefont {Hilton}, \citenamefont {Lalumière}, \citenamefont {Blais},\
  and\ \citenamefont {Lehnert}}]{chapman_widely_2017}%
  \BibitemOpen
  \bibfield  {author} {\bibinfo {author} {\bibfnamefont {B.~J.}\ \bibnamefont
  {Chapman}}, \bibinfo {author} {\bibfnamefont {E.~I.}\ \bibnamefont
  {Rosenthal}}, \bibinfo {author} {\bibfnamefont {J.}~\bibnamefont
  {Kerckhoff}}, \bibinfo {author} {\bibfnamefont {B.~A.}\ \bibnamefont
  {Moores}}, \bibinfo {author} {\bibfnamefont {L.~R.}\ \bibnamefont {Vale}},
  \bibinfo {author} {\bibfnamefont {J.~A.~B.}\ \bibnamefont {Mates}}, \bibinfo
  {author} {\bibfnamefont {G.~C.}\ \bibnamefont {Hilton}}, \bibinfo {author}
  {\bibfnamefont {K.}~\bibnamefont {Lalumière}}, \bibinfo {author}
  {\bibfnamefont {A.}~\bibnamefont {Blais}}, \ and\ \bibinfo {author}
  {\bibfnamefont {K.~W.}\ \bibnamefont {Lehnert}},\ }\href {\doibase
  10.1103/PhysRevX.7.041043} {\bibfield  {journal} {\bibinfo  {journal}
  {Physical Review X}\ }\textbf {\bibinfo {volume} {7}},\ \bibinfo {pages}
  {041043} (\bibinfo {year} {2017})}\BibitemShut {NoStop}%
\bibitem [{\citenamefont {Lecocq}\ \emph {et~al.}(2017)\citenamefont {Lecocq},
  \citenamefont {Ranzani}, \citenamefont {Peterson}, \citenamefont {Cicak},
  \citenamefont {Simmonds}, \citenamefont {Teufel},\ and\ \citenamefont
  {Aumentado}}]{lecocq_nonreciprocal_2017}%
  \BibitemOpen
  \bibfield  {author} {\bibinfo {author} {\bibfnamefont {F.}~\bibnamefont
  {Lecocq}}, \bibinfo {author} {\bibfnamefont {L.}~\bibnamefont {Ranzani}},
  \bibinfo {author} {\bibfnamefont {G.~A.}\ \bibnamefont {Peterson}}, \bibinfo
  {author} {\bibfnamefont {K.}~\bibnamefont {Cicak}}, \bibinfo {author}
  {\bibfnamefont {R.~W.}\ \bibnamefont {Simmonds}}, \bibinfo {author}
  {\bibfnamefont {J.~D.}\ \bibnamefont {Teufel}}, \ and\ \bibinfo {author}
  {\bibfnamefont {J.}~\bibnamefont {Aumentado}},\ }\href {\doibase
  10.1103/PhysRevApplied.7.024028} {\bibfield  {journal} {\bibinfo  {journal}
  {Physical Review Applied}\ }\textbf {\bibinfo {volume} {7}},\ \bibinfo
  {pages} {024028} (\bibinfo {year} {2017})}\BibitemShut {NoStop}%
\bibitem [{\citenamefont {Sliwa}\ \emph {et~al.}(2015)\citenamefont {Sliwa},
  \citenamefont {Hatridge}, \citenamefont {Narla}, \citenamefont {Shankar},
  \citenamefont {Frunzio}, \citenamefont {Schoelkopf},\ and\ \citenamefont
  {Devoret}}]{sliwa_reconfigurable_2015}%
  \BibitemOpen
  \bibfield  {author} {\bibinfo {author} {\bibfnamefont {K.~M.}\ \bibnamefont
  {Sliwa}}, \bibinfo {author} {\bibfnamefont {M.}~\bibnamefont {Hatridge}},
  \bibinfo {author} {\bibfnamefont {A.}~\bibnamefont {Narla}}, \bibinfo
  {author} {\bibfnamefont {S.}~\bibnamefont {Shankar}}, \bibinfo {author}
  {\bibfnamefont {L.}~\bibnamefont {Frunzio}}, \bibinfo {author} {\bibfnamefont
  {R.~J.}\ \bibnamefont {Schoelkopf}}, \ and\ \bibinfo {author} {\bibfnamefont
  {M.~H.}\ \bibnamefont {Devoret}},\ }\href {\doibase
  10.1103/PhysRevX.5.041020} {\bibfield  {journal} {\bibinfo  {journal}
  {Physical Review X}\ }\textbf {\bibinfo {volume} {5}},\ \bibinfo {pages}
  {041020} (\bibinfo {year} {2015})}\BibitemShut {NoStop}%
\bibitem [{\citenamefont {Ruesink}\ \emph {et~al.}(2016)\citenamefont
  {Ruesink}, \citenamefont {Miri}, \citenamefont {Alù},\ and\ \citenamefont
  {Verhagen}}]{ruesink_nonreciprocity_2016}%
  \BibitemOpen
  \bibfield  {author} {\bibinfo {author} {\bibfnamefont {F.}~\bibnamefont
  {Ruesink}}, \bibinfo {author} {\bibfnamefont {M.-A.}\ \bibnamefont {Miri}},
  \bibinfo {author} {\bibfnamefont {A.}~\bibnamefont {Alù}}, \ and\ \bibinfo
  {author} {\bibfnamefont {E.}~\bibnamefont {Verhagen}},\ }\href {\doibase
  10.1038/ncomms13662} {\bibfield  {journal} {\bibinfo  {journal} {Nature
  Communications}\ }\textbf {\bibinfo {volume} {7}},\ \bibinfo {pages} {13662}
  (\bibinfo {year} {2016})}\BibitemShut {NoStop}%
\bibitem [{\citenamefont {Fang}\ \emph {et~al.}(2017)\citenamefont {Fang},
  \citenamefont {Luo}, \citenamefont {Metelmann}, \citenamefont {Matheny},
  \citenamefont {Marquardt}, \citenamefont {Clerk},\ and\ \citenamefont
  {Painter}}]{fang_generalized_2017}%
  \BibitemOpen
  \bibfield  {author} {\bibinfo {author} {\bibfnamefont {K.}~\bibnamefont
  {Fang}}, \bibinfo {author} {\bibfnamefont {J.}~\bibnamefont {Luo}}, \bibinfo
  {author} {\bibfnamefont {A.}~\bibnamefont {Metelmann}}, \bibinfo {author}
  {\bibfnamefont {M.~H.}\ \bibnamefont {Matheny}}, \bibinfo {author}
  {\bibfnamefont {F.}~\bibnamefont {Marquardt}}, \bibinfo {author}
  {\bibfnamefont {A.~A.}\ \bibnamefont {Clerk}}, \ and\ \bibinfo {author}
  {\bibfnamefont {O.}~\bibnamefont {Painter}},\ }\href {\doibase
  10.1038/nphys4009} {\bibfield  {journal} {\bibinfo  {journal} {Nature
  Physics}\ }\textbf {\bibinfo {volume} {13}},\ \bibinfo {pages} {465}
  (\bibinfo {year} {2017})}\BibitemShut {NoStop}%
\bibitem [{\citenamefont {Wang}\ \emph {et~al.}(2019)\citenamefont {Wang},
  \citenamefont {Rao}, \citenamefont {Yang}, \citenamefont {Xu}, \citenamefont
  {Gui}, \citenamefont {Yao}, \citenamefont {You},\ and\ \citenamefont
  {Hu}}]{wang_nonreciprocity_2019}%
  \BibitemOpen
  \bibfield  {author} {\bibinfo {author} {\bibfnamefont {Y.-P.}\ \bibnamefont
  {Wang}}, \bibinfo {author} {\bibfnamefont {J.~W.}\ \bibnamefont {Rao}},
  \bibinfo {author} {\bibfnamefont {Y.}~\bibnamefont {Yang}}, \bibinfo {author}
  {\bibfnamefont {P.-C.}\ \bibnamefont {Xu}}, \bibinfo {author} {\bibfnamefont
  {Y.~S.}\ \bibnamefont {Gui}}, \bibinfo {author} {\bibfnamefont {B.~M.}\
  \bibnamefont {Yao}}, \bibinfo {author} {\bibfnamefont {J.~Q.}\ \bibnamefont
  {You}}, \ and\ \bibinfo {author} {\bibfnamefont {C.-M.}\ \bibnamefont {Hu}},\
  }\href {\doibase 10.1103/PhysRevLett.123.127202} {\bibfield  {journal}
  {\bibinfo  {journal} {Physical Review Letters}\ }\textbf {\bibinfo {volume}
  {123}},\ \bibinfo {pages} {127202} (\bibinfo {year} {2019})}\BibitemShut
  {NoStop}%
\bibitem [{\citenamefont {Xu}\ \emph {et~al.}(2019)\citenamefont {Xu},
  \citenamefont {Jiang}, \citenamefont {Clerk},\ and\ \citenamefont
  {Harris}}]{xu_nonreciprocal_2019}%
  \BibitemOpen
  \bibfield  {author} {\bibinfo {author} {\bibfnamefont {H.}~\bibnamefont
  {Xu}}, \bibinfo {author} {\bibfnamefont {L.}~\bibnamefont {Jiang}}, \bibinfo
  {author} {\bibfnamefont {A.~A.}\ \bibnamefont {Clerk}}, \ and\ \bibinfo
  {author} {\bibfnamefont {J.~G.~E.}\ \bibnamefont {Harris}},\ }\href {\doibase
  10.1038/s41586-019-1061-2} {\bibfield  {journal} {\bibinfo  {journal}
  {Nature}\ }\textbf {\bibinfo {volume} {568}},\ \bibinfo {pages} {65}
  (\bibinfo {year} {2019})}\BibitemShut {NoStop}%
\bibitem [{\citenamefont {Chen}\ \emph {et~al.}(1991)\citenamefont {Chen},
  \citenamefont {Brug},\ and\ \citenamefont
  {Goldfarb}}]{chen_demagnetizing_1991}%
  \BibitemOpen
  \bibfield  {author} {\bibinfo {author} {\bibfnamefont {D.-X.}\ \bibnamefont
  {Chen}}, \bibinfo {author} {\bibfnamefont {J.}~\bibnamefont {Brug}}, \ and\
  \bibinfo {author} {\bibfnamefont {R.}~\bibnamefont {Goldfarb}},\ }\href
  {\doibase 10.1109/20.102932} {\bibfield  {journal} {\bibinfo  {journal} {IEEE
  Transactions on Magnetics}\ }\textbf {\bibinfo {volume} {27}},\ \bibinfo
  {pages} {3601} (\bibinfo {year} {1991})}\BibitemShut {NoStop}%
\bibitem [{\citenamefont {Fletcher}\ \emph {et~al.}(1960)\citenamefont
  {Fletcher}, \citenamefont {LeCraw},\ and\ \citenamefont
  {Spencer}}]{fletcher_electron_1960}%
  \BibitemOpen
  \bibfield  {author} {\bibinfo {author} {\bibfnamefont {R.~C.}\ \bibnamefont
  {Fletcher}}, \bibinfo {author} {\bibfnamefont {R.~C.}\ \bibnamefont
  {LeCraw}}, \ and\ \bibinfo {author} {\bibfnamefont {E.~G.}\ \bibnamefont
  {Spencer}},\ }\href {\doibase 10.1103/PhysRev.117.955} {\bibfield  {journal}
  {\bibinfo  {journal} {Physical Review}\ }\textbf {\bibinfo {volume} {117}},\
  \bibinfo {pages} {955} (\bibinfo {year} {1960})}\BibitemShut {NoStop}%
\bibitem [{\citenamefont {Klingler}\ \emph {et~al.}(2017)\citenamefont
  {Klingler}, \citenamefont {Maier-Flaig}, \citenamefont {Dubs}, \citenamefont
  {Surzhenko}, \citenamefont {Gross}, \citenamefont {Huebl}, \citenamefont
  {Goennenwein},\ and\ \citenamefont {Weiler}}]{klingler_gilbert_2017}%
  \BibitemOpen
  \bibfield  {author} {\bibinfo {author} {\bibfnamefont {S.}~\bibnamefont
  {Klingler}}, \bibinfo {author} {\bibfnamefont {H.}~\bibnamefont
  {Maier-Flaig}}, \bibinfo {author} {\bibfnamefont {C.}~\bibnamefont {Dubs}},
  \bibinfo {author} {\bibfnamefont {O.}~\bibnamefont {Surzhenko}}, \bibinfo
  {author} {\bibfnamefont {R.}~\bibnamefont {Gross}}, \bibinfo {author}
  {\bibfnamefont {H.}~\bibnamefont {Huebl}}, \bibinfo {author} {\bibfnamefont
  {S.~T.~B.}\ \bibnamefont {Goennenwein}}, \ and\ \bibinfo {author}
  {\bibfnamefont {M.}~\bibnamefont {Weiler}},\ }\href {\doibase
  10.1063/1.4977423} {\bibfield  {journal} {\bibinfo  {journal} {Applied
  Physics Letters}\ }\textbf {\bibinfo {volume} {110}},\ \bibinfo {pages}
  {092409} (\bibinfo {year} {2017})}\BibitemShut {NoStop}%
\bibitem [{\citenamefont {Marković}\ \emph {et~al.}(2018)\citenamefont
  {Marković}, \citenamefont {Jezouin}, \citenamefont {Ficheux}, \citenamefont
  {Fedortchenko}, \citenamefont {Felicetti}, \citenamefont {Coudreau},
  \citenamefont {Milman}, \citenamefont {Leghtas},\ and\ \citenamefont
  {Huard}}]{markovic_demonstration_2018}%
  \BibitemOpen
  \bibfield  {author} {\bibinfo {author} {\bibfnamefont {D.}~\bibnamefont
  {Marković}}, \bibinfo {author} {\bibfnamefont {S.}~\bibnamefont {Jezouin}},
  \bibinfo {author} {\bibfnamefont {Q.}~\bibnamefont {Ficheux}}, \bibinfo
  {author} {\bibfnamefont {S.}~\bibnamefont {Fedortchenko}}, \bibinfo {author}
  {\bibfnamefont {S.}~\bibnamefont {Felicetti}}, \bibinfo {author}
  {\bibfnamefont {T.}~\bibnamefont {Coudreau}}, \bibinfo {author}
  {\bibfnamefont {P.}~\bibnamefont {Milman}}, \bibinfo {author} {\bibfnamefont
  {Z.}~\bibnamefont {Leghtas}}, \ and\ \bibinfo {author} {\bibfnamefont
  {B.}~\bibnamefont {Huard}},\ }\href {\doibase 10.1103/PhysRevLett.121.040505}
  {\bibfield  {journal} {\bibinfo  {journal} {Physical Review Letters}\
  }\textbf {\bibinfo {volume} {121}},\ \bibinfo {pages} {040505} (\bibinfo
  {year} {2018})}\BibitemShut {NoStop}%
\bibitem [{\citenamefont {Solt}(1962)}]{solt_temperature_1962}%
  \BibitemOpen
  \bibfield  {author} {\bibinfo {author} {\bibfnamefont {I.~H.}\ \bibnamefont
  {Solt}},\ }\href {\doibase 10.1063/1.1728651} {\bibfield  {journal} {\bibinfo
   {journal} {Journal of Applied Physics}\ }\textbf {\bibinfo {volume} {33}},\
  \bibinfo {pages} {1189} (\bibinfo {year} {1962})}\BibitemShut {NoStop}%
\bibitem [{\citenamefont {Schlömann}(1970)}]{schlomann_microwave_1970}%
  \BibitemOpen
  \bibfield  {author} {\bibinfo {author} {\bibfnamefont {E.}~\bibnamefont
  {Schlömann}},\ }\href {\doibase 10.1063/1.1658322} {\bibfield  {journal}
  {\bibinfo  {journal} {Journal of Applied Physics}\ }\textbf {\bibinfo
  {volume} {41}},\ \bibinfo {pages} {204} (\bibinfo {year} {1970})}\BibitemShut
  {NoStop}%
\bibitem [{\citenamefont {Green}\ and\ \citenamefont
  {Sandy}(1974)}]{green_microwave_1974}%
  \BibitemOpen
  \bibfield  {author} {\bibinfo {author} {\bibfnamefont {J.}~\bibnamefont
  {Green}}\ and\ \bibinfo {author} {\bibfnamefont {F.}~\bibnamefont {Sandy}},\
  }\href {\doibase 10.1109/TMTT.1974.1128306} {\bibfield  {journal} {\bibinfo
  {journal} {IEEE Transactions on Microwave Theory and Techniques}\ }\textbf
  {\bibinfo {volume} {22}},\ \bibinfo {pages} {641} (\bibinfo {year}
  {1974})}\BibitemShut {NoStop}%
\bibitem [{\citenamefont {Casimir}(1945)}]{casimir_onsagers_1945}%
  \BibitemOpen
  \bibfield  {author} {\bibinfo {author} {\bibfnamefont {H.~B.~G.}\
  \bibnamefont {Casimir}},\ }\href {\doibase 10.1103/RevModPhys.17.343}
  {\bibfield  {journal} {\bibinfo  {journal} {Reviews of Modern Physics}\
  }\textbf {\bibinfo {volume} {17}},\ \bibinfo {pages} {343} (\bibinfo {year}
  {1945})}\BibitemShut {NoStop}%
\bibitem [{\citenamefont {Ranzani}\ \emph {et~al.}(2013)\citenamefont
  {Ranzani}, \citenamefont {Spietz}, \citenamefont {Popovic},\ and\
  \citenamefont {Aumentado}}]{ranzani_two-port_2013}%
  \BibitemOpen
  \bibfield  {author} {\bibinfo {author} {\bibfnamefont {L.}~\bibnamefont
  {Ranzani}}, \bibinfo {author} {\bibfnamefont {L.}~\bibnamefont {Spietz}},
  \bibinfo {author} {\bibfnamefont {Z.}~\bibnamefont {Popovic}}, \ and\
  \bibinfo {author} {\bibfnamefont {J.}~\bibnamefont {Aumentado}},\ }\href
  {\doibase 10.1063/1.4794910} {\bibfield  {journal} {\bibinfo  {journal}
  {Review of Scientific Instruments}\ }\textbf {\bibinfo {volume} {84}},\
  \bibinfo {pages} {034704} (\bibinfo {year} {2013})}\BibitemShut {NoStop}%
\bibitem [{\citenamefont {Helszajn}(2008)}]{helszajn_stripline_2008}%
  \BibitemOpen
  \bibfield  {author} {\bibinfo {author} {\bibfnamefont {J.}~\bibnamefont
  {Helszajn}},\ }\href@noop {} {\emph {\bibinfo {title} {The {Stripline}
  {Circulator}: {Theory} and {Practice}}}},\ \bibinfo {edition} {1st}\ ed.\
  (\bibinfo  {publisher} {Wiley-IEEE Press},\ \bibinfo {address} {Hoboken,
  NJ},\ \bibinfo {year} {2008})\BibitemShut {NoStop}%
\bibitem [{\citenamefont {Metelmann}\ and\ \citenamefont
  {Clerk}(2015)}]{metelmann_nonreciprocal_2015}%
  \BibitemOpen
  \bibfield  {author} {\bibinfo {author} {\bibfnamefont {A.}~\bibnamefont
  {Metelmann}}\ and\ \bibinfo {author} {\bibfnamefont {A.~A.}\ \bibnamefont
  {Clerk}},\ }\href {\doibase 10.1103/PhysRevX.5.021025} {\bibfield  {journal}
  {\bibinfo  {journal} {Physical Review X}\ }\textbf {\bibinfo {volume} {5}},\
  \bibinfo {pages} {021025} (\bibinfo {year} {2015})}\BibitemShut {NoStop}%
\bibitem [{\citenamefont {McDonald}\ and\ \citenamefont
  {Clerk}(2020)}]{mcdonald_exponentially-enhanced_2020}%
  \BibitemOpen
  \bibfield  {author} {\bibinfo {author} {\bibfnamefont {A.}~\bibnamefont
  {McDonald}}\ and\ \bibinfo {author} {\bibfnamefont {A.~A.}\ \bibnamefont
  {Clerk}},\ }\href {\doibase 10.1038/s41467-020-19090-4} {\bibfield  {journal}
  {\bibinfo  {journal} {Nature Communications}\ }\textbf {\bibinfo {volume}
  {11}},\ \bibinfo {pages} {5382} (\bibinfo {year} {2020})}\BibitemShut
  {NoStop}%
\bibitem [{\citenamefont {Schomerus}(2020)}]{schomerus_nonreciprocal_2020}%
  \BibitemOpen
  \bibfield  {author} {\bibinfo {author} {\bibfnamefont {H.}~\bibnamefont
  {Schomerus}},\ }\href {\doibase 10.1103/PhysRevResearch.2.013058} {\bibfield
  {journal} {\bibinfo  {journal} {Physical Review Research}\ }\textbf {\bibinfo
  {volume} {2}},\ \bibinfo {pages} {013058} (\bibinfo {year}
  {2020})}\BibitemShut {NoStop}%
\bibitem [{\citenamefont {Sounas}\ and\ \citenamefont
  {Alù}(2017)}]{sounas_non-reciprocal_2017}%
  \BibitemOpen
  \bibfield  {author} {\bibinfo {author} {\bibfnamefont {D.~L.}\ \bibnamefont
  {Sounas}}\ and\ \bibinfo {author} {\bibfnamefont {A.}~\bibnamefont {Alù}},\
  }\href {\doibase 10.1038/s41566-017-0051-x} {\bibfield  {journal} {\bibinfo
  {journal} {Nature Photonics}\ }\textbf {\bibinfo {volume} {11}},\ \bibinfo
  {pages} {774} (\bibinfo {year} {2017})}\BibitemShut {NoStop}%
\bibitem [{\citenamefont {Ramos}\ \emph {et~al.}(2014)\citenamefont {Ramos},
  \citenamefont {Pichler}, \citenamefont {Daley},\ and\ \citenamefont
  {Zoller}}]{ramos_quantum_2014}%
  \BibitemOpen
  \bibfield  {author} {\bibinfo {author} {\bibfnamefont {T.}~\bibnamefont
  {Ramos}}, \bibinfo {author} {\bibfnamefont {H.}~\bibnamefont {Pichler}},
  \bibinfo {author} {\bibfnamefont {A.~J.}\ \bibnamefont {Daley}}, \ and\
  \bibinfo {author} {\bibfnamefont {P.}~\bibnamefont {Zoller}},\ }\href
  {\doibase 10.1103/PhysRevLett.113.237203} {\bibfield  {journal} {\bibinfo
  {journal} {Physical Review Letters}\ }\textbf {\bibinfo {volume} {113}},\
  \bibinfo {pages} {237203} (\bibinfo {year} {2014})}\BibitemShut {NoStop}%
\bibitem [{\citenamefont {Guimond}\ \emph {et~al.}(2020)\citenamefont
  {Guimond}, \citenamefont {Vermersch}, \citenamefont {Juan}, \citenamefont
  {Sharafiev}, \citenamefont {Kirchmair},\ and\ \citenamefont
  {Zoller}}]{guimond_unidirectional_2020}%
  \BibitemOpen
  \bibfield  {author} {\bibinfo {author} {\bibfnamefont {P.-O.}\ \bibnamefont
  {Guimond}}, \bibinfo {author} {\bibfnamefont {B.}~\bibnamefont {Vermersch}},
  \bibinfo {author} {\bibfnamefont {M.~L.}\ \bibnamefont {Juan}}, \bibinfo
  {author} {\bibfnamefont {A.}~\bibnamefont {Sharafiev}}, \bibinfo {author}
  {\bibfnamefont {G.}~\bibnamefont {Kirchmair}}, \ and\ \bibinfo {author}
  {\bibfnamefont {P.}~\bibnamefont {Zoller}},\ }\href {\doibase
  10.1038/s41534-020-0261-9} {\bibfield  {journal} {\bibinfo  {journal} {npj
  Quantum Information}\ }\textbf {\bibinfo {volume} {6}},\ \bibinfo {pages} {1}
  (\bibinfo {year} {2020})}\BibitemShut {NoStop}%
\bibitem [{\citenamefont {Gheeraert}\ \emph {et~al.}(2020)\citenamefont
  {Gheeraert}, \citenamefont {Kono},\ and\ \citenamefont
  {Nakamura}}]{gheeraert_programmable_2020}%
  \BibitemOpen
  \bibfield  {author} {\bibinfo {author} {\bibfnamefont {N.}~\bibnamefont
  {Gheeraert}}, \bibinfo {author} {\bibfnamefont {S.}~\bibnamefont {Kono}}, \
  and\ \bibinfo {author} {\bibfnamefont {Y.}~\bibnamefont {Nakamura}},\ }\href
  {\doibase 10.1103/PhysRevA.102.053720} {\bibfield  {journal} {\bibinfo
  {journal} {Physical Review A}\ }\textbf {\bibinfo {volume} {102}},\ \bibinfo
  {pages} {053720} (\bibinfo {year} {2020})}\BibitemShut {NoStop}%
\bibitem [{\citenamefont {Rymarz}\ \emph {et~al.}(2021)\citenamefont {Rymarz},
  \citenamefont {Bosco}, \citenamefont {Ciani},\ and\ \citenamefont
  {DiVincenzo}}]{rymarz_hardware-encoding_2021}%
  \BibitemOpen
  \bibfield  {author} {\bibinfo {author} {\bibfnamefont {M.}~\bibnamefont
  {Rymarz}}, \bibinfo {author} {\bibfnamefont {S.}~\bibnamefont {Bosco}},
  \bibinfo {author} {\bibfnamefont {A.}~\bibnamefont {Ciani}}, \ and\ \bibinfo
  {author} {\bibfnamefont {D.~P.}\ \bibnamefont {DiVincenzo}},\ }\href
  {\doibase 10.1103/PhysRevX.11.011032} {\bibfield  {journal} {\bibinfo
  {journal} {Physical Review X}\ }\textbf {\bibinfo {volume} {11}},\ \bibinfo
  {pages} {011032} (\bibinfo {year} {2021})}\BibitemShut {NoStop}%
\end{thebibliography}%

\end{document}